\documentclass[a4paper,11pt]{article}
\usepackage{jheppub} 
\usepackage{lineno}
\usepackage{graphicx} 
\usepackage{amsmath}
\usepackage{tikz}
\usepackage{caption}
\usepackage{subcaption}
\usepackage{amssymb}
\usepackage{braket}
\usepackage{enumitem}

\title{\boldmath Gravity Dual of Networks}

\author{Yu Guo, Rong-Xin Miao}
\affiliation{School of Physics and Astronomy, Sun Yat-Sen University,\\
2 Daxue Road, Zhuhai 519082, China}

\emailAdd{guoy225@mail2.sysu.edu.cn}
\emailAdd{miaorx@mail.sysu.edu.cn}

\abstract{The network has been attracting increasing attention for its role in driving the artificial intelligence revolution and enabling profound insights into gravity. This paper investigates the gravity dual of the conformal field theory defined on a network (AdS/NCFT). A typical network, consisting of edges and nodes, is dual to a spacetime with branches and connecting branes, which we refer to as Net-branes. We demonstrate that the junction condition on the Net-brane results in energy conservation at the network node, providing strong support for our proposal of AdS/NCFT. 

We find that the spectrum of gravitational Kaluza-Klein modes on the Net-brane is a combination of the spectra from the AdS/BCFT with Neumann boundary conditions and Dirichlet/Conformal boundary conditions, corresponding to the isolated and transparent modes, respectively. We study two-point functions for NCFTs and provide examples, such as free fields and AdS/NCFT with tensionless Net-branes. We propose that the RT surfaces intersect at the same point on the Net-brane for connected subsystems within the network and verify this with the strong additivity and monotonicity of entanglement entropy. We establish that the network entropy, defined as the difference in entanglement between NCFT and BCFT, is always non-negative and effectively illustrates the network's complexity. Finally, we briefly discuss the holographic perspective of the shortest path problem and reveal its relation to the shortest geodesic in bulk and the holographic two-point correlators of massive operators.}

\begin{document}

\maketitle

\flushbottom

\section{Introduction}

Everything in the universe is interconnected, whether through interactions, quantum correlations, or entanglement. Networks offer a strong framework for studying these connections and play a crucial role in physics. Notably, neural networks are driving the recent revolution in artificial intelligence \cite{Hopfield:1982pe, Rumelhart:1986gxv, Hinton:2006tev}. Additionally, tensor networks have enhanced our understanding of quantum entanglement and gravity \cite{Swingle:2009bg, Hayden:2016cfa, Chen:2021ipv}. Many physical systems, such as complex circuits, multi-node optical fibers, and microfluidic channels, naturally exhibit network structures. Therefore, studying physics within the context of networks is of significant theoretical and practical importance.

On the other hand, holography provides a deep insight into our understanding of gravity \cite{tHooft:1993dmi, Susskind:1994vu, Maldacena:1997re}. It proposes that quantum gravity in bulk is dual to a conformal field theory (CFT) on the boundary \cite{Maldacena:1997re}. It reveals a profound connection between spacetime geometry and quantum entanglement \cite{Ryu:2006bv}. It provides a robust framework for studying strongly coupled systems and predicts the existence of a lower bound for fluid shear viscosity ratio entropy density \cite{Kovtun:2004de}. Naturally, this raises the question: Can holography offer new insights into the physics of networks? For instance, what would the holographic dual of shortest-path problems be? And how might an efficient neural network manifest from a gravitational duality perspective? 
 \begin{figure}[htbp]
  \centering
\includegraphics[width=.3\textwidth]{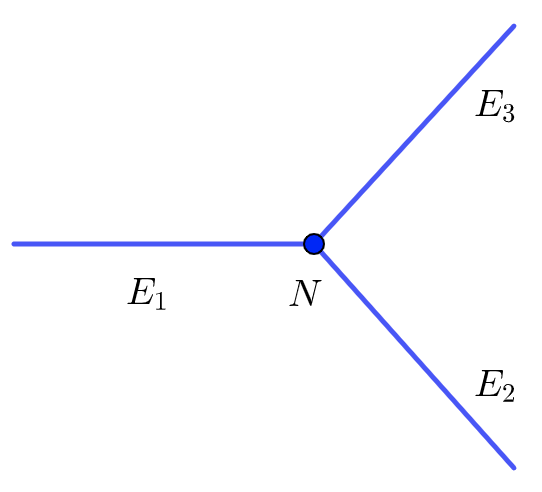}
\qquad \qquad
\includegraphics[width=.4\textwidth]{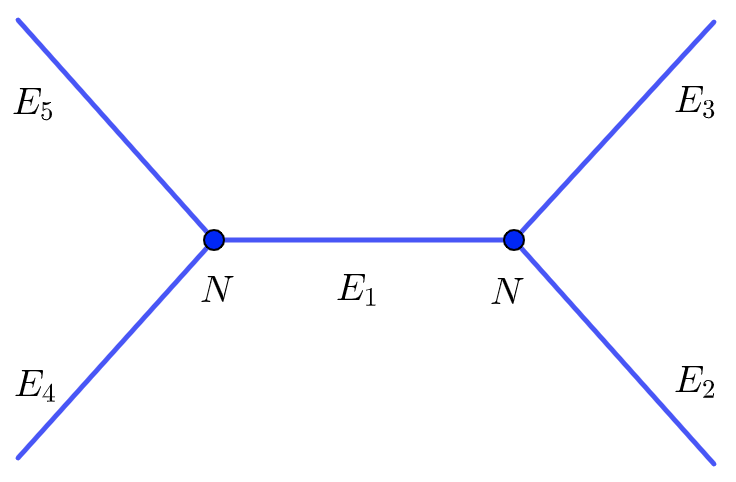}
 \caption{The networks with one node (left) and two nodes (right). The blue lines and points denote the edges ($E_m$) and nodes ($N$) of the networks.  } 
 \label{network}
\end{figure}

This paper studies the CFT on networks (NCFT) and its gravity dual. A typical network includes several edges and nodes. See Fig. \ref{network} for some examples. The nodes link the CFTs living on different edges. Thus, a key step in developing the theory of NCFT is determining the connecting conditions on the nodes. To do so, let us compare NCFT with the CFT with a boundary (BCFT). The reflective boundary condition for a BCFT reads 
\begin{align}\label{BCFT BC}
\text{BCFT}: \ J_n|_{\text{bdy}}=0,\ \  \ \  T_{na}|_{\text{bdy}}=0,
\end{align}
which means no current and energy can flow out of the boundary. Here, $n, a$ label the normal and tangential directions, and $J$ and $T$ denote the current and stress tensor. For an NCFT, the current and energy conservations suggest
\begin{align}\label{NCFT BC}
\text{NCFT}: \ \sum_{m} \overset{(m)}{J}_n|_{\text{node}}=0, \ \  \sum_{m} \overset{(m)}{T}_{na}|_{\text{node}}=0,
\end{align}
which means the total current and energy/tangential momentum flowing into a node is zero. Here, $ \overset{(m)}{J}$ and $ \overset{(m)}{T}$are the current and stress tensor on the edge $E_m$ linked by the same node. Note that since the current/energy can flow from one edge to another, $\overset{(m)}{J}_n|_{\text{node}}$ and $\overset{(m)}{T}_{na}|_{\text{node}}$ on edge $E_m$ can be non-zero, which is the main difference from a BCFT. From the stress tensor, we can construct a current $J_i=T_{ij} \xi^j$, where $\xi^j=\delta^j_a$ are Killing vectors tangential to the node. Note that the translation invariance normal to the node is broken, so $x^j=\delta^j_n$ is no longer a Killing vector. Then the first equation of (\ref{NCFT BC}) implies the second one. It's important to note that there are no constraints on the tangential components of the current or the tangential-tangential components of the stress tensor at the node. In particular, these components are generally not continuous across the node. To illustrate this, let's consider the interface in Maxwell's theory. Consider two conductor layers with different conductivities $\sigma$, Ohm's law ${\bf J}=\sigma {\bf E}$, along with the continuity of the tangential electric field $E_{1a}=E_{2a}=E_{a}$, typically results in discontinuous tangential currents $J_{1a}/\sigma_1=J_{2a}/\sigma_2=E_{a}$ across the interface. Now, consider two dielectric layers with different permittivities \(\epsilon\) and a tangential-only electric field \(E_a\). The corresponding energy densities at the interface are discontinuous for $\epsilon_1\ne \epsilon_2$: \(T_{1\ tt}/\epsilon_1 =T_{2\ tt}/\epsilon_2 = E_{a}^2/2 \).

 \begin{figure}[htbp]
  \centering
\includegraphics[width=.3\textwidth]{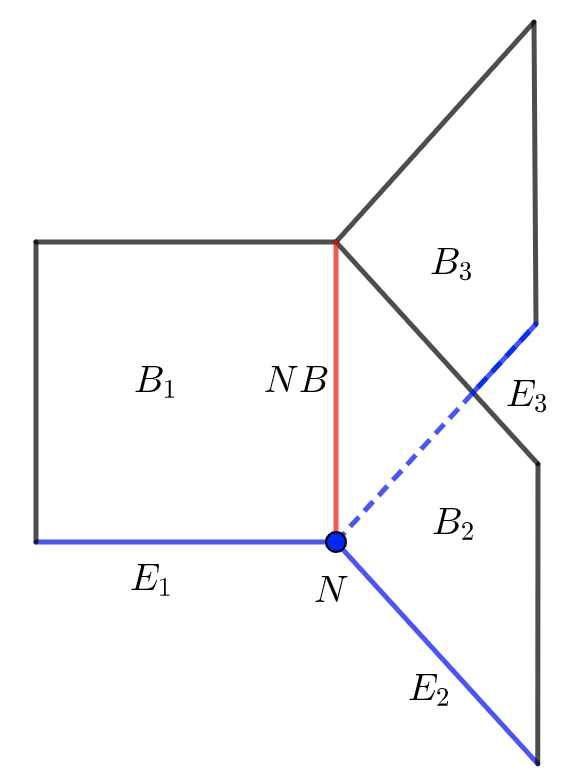}
\qquad \qquad
\includegraphics[width=.4\textwidth]{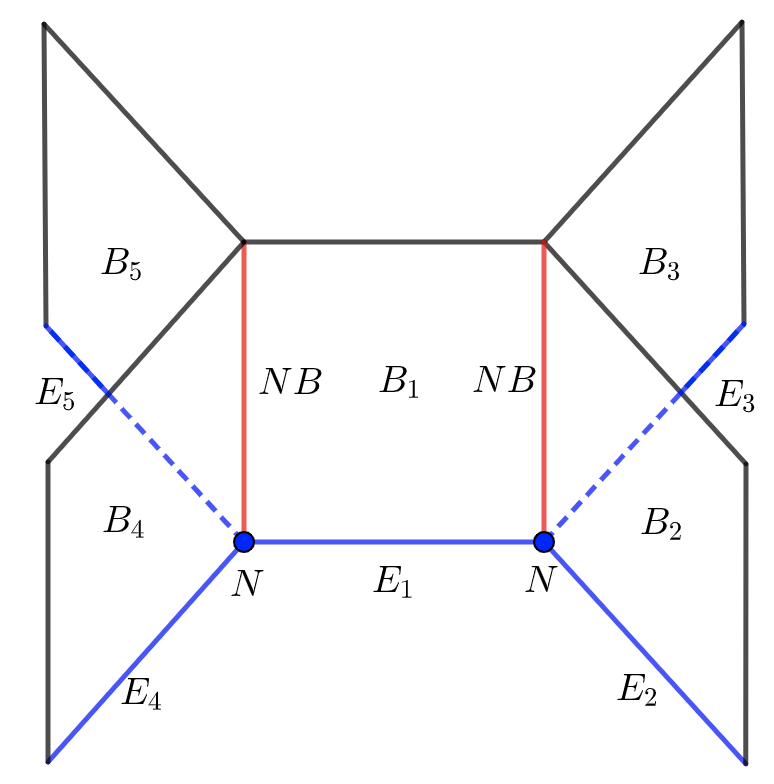}
 \caption{Geometries for holographic networks. The blue lines and points denote the edges ($E_m$) and nodes ($N$) of the networks. The red lines label the Net-branes $NB$, which link the branches $B_m$ (squares) in bulk. The edges ($E_m$) and nodes ($N$) are dual to the branches $B_m$ and Net-branes $NB$ in bulk, respectively. For simplicity, we show only the holographic duals of the networks of Fig. \ref{network}. One can glue above geometries to get the gravity duals of general networks. } \label{holo network}
\end{figure}

Let us go on to discuss the gravity dual. The boundary of a BCFT is dual to an end-of-the-world (EOW) brane in bulk. Takayanagi proposes to impose Neumann boundary condition (NBC) on the EOW brane \cite{Takayanagi:2011zk}, 
\begin{align}\label{NBC BCFT}
\text{AdS/BCFT}:\ \Big(K_{ij}-K h_{ij}\Big)|_{\text{EOW brane}}=-T h_{ij},
\end{align}
which results in (\ref{BCFT BC}) for the dual BCFTs \footnote{To the best of our knowledge, there is significant evidence supporting this assertion, but no rigorous proof has been established yet. We prove this statement for AdS/BCFT and its variant for AdS/NCFT in this paper.}. Here, $K_{ij}$ are extrinsic curvatures, $h_{ij}$ are induced metrics, and $T$ is the brane tension. Similarly, the network node is dual to a brane in bulk, and we refer to it as a Net-brane. See Fig. \ref{holo network} for the geometry of holographic networks. Inspired by (\ref{NBC BCFT}), we propose to impose the following junction condition on the Net-brane
\begin{align}\label{junction condition NCFT}
\text{AdS/NCFT}:\ \sum_m \Big(\overset{(m)}{K}_{ij}-\overset{(m)}{K} h_{ij}\Big)|_{\text{Net-brane}}=-T h_{ij},
\end{align}
where $\overset{(m)}{K}_{ij}$ denote the extrinsic curvatures from bulk branches $B_m$ to the Net-brane.
We require the induced metrics from different bulk branches into the Net-brane to be the same 
\begin{align}\label{continuity condition NCFT}
\text{AdS/NCFT}:\  \overset{(m)}{h}_{ij}|_{\text{Net-brane}}= h_{ij}.
\end{align}
Recall that the NBC (\ref{NBC BCFT}) in bulk results in the reflective boundary condition (\ref{BCFT BC}) for BCFTs. Similarly, the bulk junction condition (\ref{junction condition NCFT}) yields the conservation law (\ref{NCFT BC}) for NCFTs on the network node. Note that  (\ref{junction condition NCFT}) reduces to the usual junction condition when there are only two bulk branches \footnote{The usual junction condition on a thin shell reads $(K^{R}_{ij}-K^R h_{ij})+(K^{L}_{ij}-K^L h_{ij})=-T h_{ij}$, where R and L denote right and left-hand sides, and $K_{ij}=h^{i_1}_i h^{j_1}_j\nabla_{i_1} n_{j_1}$ with $n_i$ the outpointing normal vector. Note that $n^R_i$ and $n^L_i$ have opposite signs in our notations.}. It provides additional support for our proposal. It should be mentioned that the junction condition (\ref{junction condition NCFT}) has recently been obtained in \cite{Shen:2024itl} for spacetimes with multiple boundaries, such as wormholes. Here, we have different physical motivations, focusing on the gravity dual of NCFT. In particular, our spacetime has only one boundary (the network) instead of multiple boundaries.  

In this paper, we derive the junction condition (\ref{junction condition NCFT}) from the variational principle and prove that it leads to the conservation law (\ref{NCFT BC}) for NCFTs. We examine various geometries of AdS/NCFT and discover that the spectrum of gravitational Kaluza-Klein (KK) modes on the Net-brane is a mix of the spectra from the AdS/BCFT with NBC \cite{Takayanagi:2011zk} and DBC/CBC \cite{Miao:2018qkc, Chu:2021mvq}. Additionally, we investigate the general forms of two-point correlation functions in NCFTs. We study examples of free theories in the CFT and the tensionless Net-brane in the gravitational dual. We confirm that the current two-point functions satisfy the essential condition (\ref{NCFT BC}) at the network node. Next, we explore holographic entanglement entropy (HEE) in AdS/NCFT, which reveals interesting structures. We propose that for a connected subsystem on the network, the Ryu-Takayanagi (RT) surfaces in different bulk branches must be connected, but only on the Net-branes. Our proposal is consistent with the principles of strong additivity and monotonicity of entanglement entropy. We derive the connecting condition for RT surfaces on the Net-brane, showing it simplifies to the orthogonality condition for the RT surface and EOW brane in AdS/BCFT under $Z_p$ symmetry. We outline several natural definitions of network entropy; one is based on the difference in entanglement entropy between NCFT and BCFT, which is always non-negative and effectively illustrates the network's structure and complexity. Finally, we briefly discuss the holographic viewpoints of famous network problems, such as the shortest path problem. We find it is closely related to the shortest geodesic in bulk and the holographic two-point correlators of massive operators.  

The paper is structured as follows. In Section 2, we formulate the holographic dual of networks and demonstrate that the junction condition leads to a conservation law on the node for NCFTs. We also explore typical solutions in the AdS/NCFT framework and analyze the spectrum of gravitational KK modes on the Net-brane. Section 3 examines the general forms of two-point functions for NCFTs and provides examples, including free fields and AdS/NCFT with tensionless Net-branes. Section 4 focuses on holographic entanglement entropy. We propose that the RT surfaces intersect at the same point on the Net-brane for connected subsystems within the network and discuss potential definitions of network entropy. Section 5 addresses the holographic shortest path problem for networks, highlighting its connection to the shortest geodesic in bulk and the holographic two-point correlators of massive operators. Finally, we conclude with a discussion of open issues in Section 6. 
Appendix \ref{app for notations} provides a summary of the notations used in this paper. Appendix \ref{app for Killing vectors} examines the local Killing vectors in the linear order. 
Appendix \ref{app for holo one-point function} discusses the holographic one-point function of the scalar operator in the simplest network consisting of only one node. Appendix \ref{app for two point function} derives the holographic two-point function of scalar operators in general networks.

\section{Gravity dual of NCFT}
\label{sect for gravity dual of NCFT}

This section explores the gravity dual of NCFTs. We derive the junction condition  (\ref{junction condition NCFT}) from the variational principle in the bulk and demonstrate that it results in the conservation law (\ref{NCFT BC}) for NCFTs. We also examine typical AdS/NCFT solutions and analyze the spectrum of KK modes on the Net-brane. For simplicity, we focus on the basic network illustrated in Fig. \ref{network} (left). The generalization to more complex networks is straightforward.

\begin{figure}[htbp]
  \centering
\includegraphics[width=0.35\textwidth]{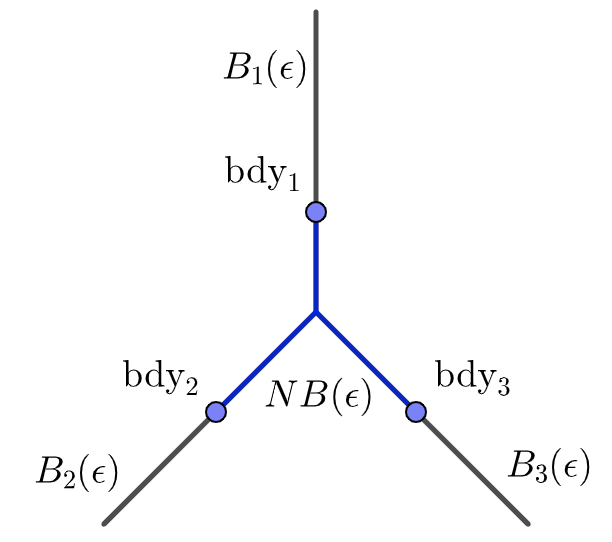}
 \caption{We regularize the Net-brane by extending $\epsilon$ into the branch $B_m$. The blue region denotes the regularized Net-brane $NB(\epsilon)$, and the blue points are the boundary of the regularized branch $B_m(\epsilon)$. In this regularization, the bulk gravity is constrained to
$B_m(\epsilon)$. For a well-defined action principle, we add Gibbons-Hawking-York terms $\overset{(m)}{K}$ on the boundary of $B_m(\epsilon)$.  }
  \label{reNetbrane}
\end{figure}

Let us start with the geometry of the holographic networks. See Fig. \ref{holo network} for examples. The simplest network comprises $p$ edges $E_m$ and one node $N$, which are dual to the branches $B_m$ and the Net-brane $NB$ in bulk, respectively. We have $\partial B_m=E_m\cup NB$ and $\partial (NB)=N$. The gravitational action in bulk reads 
\begin{align}\label{sect2: gravity action}
I=\frac{1}{16\pi G_N}\sum_m^p \int_{B_m} d^{d+1}x\sqrt{|g|} (R-2\Lambda) +\frac{1}{8\pi G_N}\int_{NB} d^dy \sqrt{|h|}(-T+\sum_m^p \overset{(m)}{K})+I_{\text{matter}}(\Psi),
\end{align}
where $-2\Lambda=d(d-1)/l^2$ is the cosmological constant, $\overset{(m)}{K}$ represents the extrinsic curvature on the Net-brane coming from the branch $B_m$, and $\Psi$ denotes various matter fields. For simplicity, we assume the matters are minimally coupled to the bulk gravity and there is no matter action except a tension term $T$ on the Net-brane. We include the Gibbons-Hawking-York (GHY) term $\overset{(m)}{K}$ on the Net-brane to ensure a well-defined action principle \footnote{As illustrated in Fig. \ref{reNetbrane}, we regularize the Net-brane by extending $\epsilon$ into the branch $B_m$. The bulk gravity is constrained to $B_m(\epsilon)$ and the brane tension is diffused in $NB(\epsilon)$. Naturally, we add GHY terms on the boundaries of $B_m$. By taking the limit $\epsilon \to$, we recover the terms on Net-brane in (\ref{sect2: gravity action}).}. We assume the induced metric from different branches $B_m$ to the Net-brane $NB$ remains continuous, i.e., $ \overset{(m)}{h}_{ij}|_{NB}= h_{ij}$. And similar for the induced matter fields, i.e., $ \overset{(m)}{\Psi}|_{NB}= \Psi$. For simplicity, we set $16\pi G_N=1$ and $l=1$ in the following discussions\footnote{In general, the Newton's constant $G_N$ and the AdS radius $l$ can take different values on different branches $B_m$. We focus on the simplest case with the same $G_N$ and $l$ on all the branches in this paper. }. Taking variations of the action (\ref{sect2: gravity action}), we get on the Net-brane 
\begin{align}\label{sect2: bdy dI}
\delta I|_{NB}=\int_{NB} d^dy \sqrt{|h|} \Big[\Big(T h_{ij}+\sum_m^p (\overset{(m)}{K}_{ij}-\overset{(m)}{K} h_{ij}) \Big)\delta h^{ij}+\sum_m^p\overset{(m)}{\Pi}\delta \Psi \Big] =0,
\end{align}
where $\overset{(m)}{\Pi}$ are the conjugate momenta of matter fields along the normal direction.  
Since the induced metric and matter fields on the Net-brane can vary, i.e., $\delta h_{ij}, \delta \Psi\neq 0$, we derive the junction conditions for gravity and matter fields
\begin{align}\label{sect2: junction condition gravity matter}
\sum_m^p \Big(\overset{(m)}{K}_{ij}-\overset{(m)}{K} h_{ij}\Big)|_{NB}=-T h_{ij},\ \ \ \ \ \sum_m^p\overset{(m)}{\Pi}|_{NB}=0. 
\end{align}
For example, when considering bulk scalars and the Maxwell field, the matter action is
\begin{align}\label{sect2: matter action}
I_{\text{matter}}=\sum_m^p \int_{B_m} d^{d+1}x\sqrt{|g|} \Big(-\frac{1}{2} \nabla_{\mu} \Phi  \nabla^{\mu} \Phi -\frac{1}{2} m^2 \Phi^2-\frac{1}{4} \mathcal{F}^{\mu\nu} \mathcal{F}_{\mu\nu}\Big).
\end{align}
Then, the second equation of (\ref{sect2: junction condition gravity matter}) can be written exactly as
\begin{align}\label{sect2: bdy dI matter}
\sum_m^p \nabla_{\hat{n}} \overset{(m)}{\Phi}|_{NB}=0,\ \   \sum_m^p \overset{(m)}{\mathcal{F}}_{\hat{n} i}|_{NB}=0,
\end{align}
where $\hat{n}, i$ are bulk normal and tangential directions to the Net-brane. Please distinguish them from $n, a$, which label the normal and tangential directions to the node on either edge $E_m$ or the Net-brane $NB$. From the junction condition (\ref{sect2: bdy dI matter}), continuity condition $\overset{(m)}{\Phi}|_{NB}= \Phi,  \overset{(m)}{\mathcal{F}}_{ij}|_{NB}= \mathcal{F}_{ij}$, and 
the expression of the normal-tangential components of the bulk stress tensor
\begin{align}\label{sect2: matter stress}
\overset{(m)}{T}{}^{\text{matter}}_{\hat{n}j}=\nabla_{\hat{n}} \overset{(m)}{\Phi} \nabla_j \overset{(m)}{\Phi} + \overset{(m)}{\mathcal{F}}_{\hat{n} i} \overset{(m)}{\mathcal{F}}{}_j^{\ i},
\end{align}
we derive 
\begin{align}\label{sect2: matter Tij JC}
\sum_{m}^p\overset{(m)}{T}{}^{\text{matter}}_{\hat{n}j}|_{NB}=0.
\end{align}
It means the total energy and tangential momentum fluxes into the Net-brane are zero, which is the bulk counterpart of the conservation law (\ref{NCFT BC}) in networks.

\subsection{Conservation law on nodes} \label{section 2.1}

Now turn to the important part of this section. We want to demonstrate how the junction condition (\ref{junction condition NCFT}) leads to the energy conservation law (\ref{NCFT BC}) at the network node. 
Here is a simplified outline. We start by constructing a conserved current \( J_i = T_{ij}\xi^j \) using the stress tensor and local Killing vectors on the edges and the Net-brane. Next, we apply Gauss's law near the node. It helps us relate the normal components of the currents on the edges and the Net-brane, resulting in $\sum_m( \overset{(m)}{J_E}_{\ n} + \overset{(m)}{J_{NB}}_{\ n})|_N=0$. This expresses the idea that the total current flowing into the node from both the edges and the Net-brane equals zero. Then, we use the junction condition on the Net-brane to show that \( \sum_m \overset{(m)}{J_{NB}}_{\ n}|_N = 0 \), which leads to the current conservation law for NCFTs: $\sum_m  \overset{(m)}{J_E}_{\ n} |_N=0$. Finally, noting that $\xi^j\sim \delta^j_a$ with $a$ the tangential directions at the node, we obtain the energy conservation law at the node:  $\sum_m \overset{(m)}{T_E}_{\ na}|_N=0$.  

We start with Codazzi's equation:
\begin{align}\label{sect2: C equation}
D^i (2K h_{ij}-2K_{ij}+t_{ij})=-2R_{\hat{n}j}=-T^{\text{matter}}_{\hat{n}j},
\end{align}
where $D_i$ represents the covariant derivative, $K_{ij}$ is the extrinsic curvature, and $t_{ij}$ is a conserved stress tensor on the boundary, satisfying $D^it_{ij}=0$. We will specify $t_{ij}$ later. Here, $R_{\hat{n}j}$ and  $T^{\text{matter}}_{\hat{n}j}$ refer to the Ricci tensor and matter stress tensor in the bulk, respectively, with $\hat{n}$ indicating the normal direction and $j$ the tangential direction to the boundary. On the AdS boundary, $T^{\text{matter}}_{\hat{n}j}$ vanishes, meaning no matter flows out of the boundary. On the Net-brane, as shown in (\ref{sect2: matter Tij JC}), only the sum $\sum_{m}^p\overset{(m)}{T}{}^{\text{matter}}_{\hat{n}j}|_{NB}=0$ vanishes. It implies that the original Codazzi equation vanishes, and we find that $T_{ij} = 2K h_{ij} - 2K_{ij} + t_{ij}$ represents conserved energy-momentum tensors on both the AdS boundary $E_m$ and the Net-brane $NB$ (in the sense of sum) 
\begin{align}\label{sect2: C equation Bm}
&D^i (2K h_{ij}-2K_{ij}+t_{ij})|_{E_m}=0, \\
& \sum_{m} D^i (2\overset{(m)}{K} h_{ij}-2\overset{(m)}{K}_{ij}+\overset{(m)}{t}_{ij})|_{NB}=0. \label{sect2: C equation NB}
\end{align}
By adding the Hayward term \cite{Hayward:1993my}, we can smooth the above equations on the intersection of $E_m$ and $NB$, i.e., $N=E_m\cap NB$. A natural way is to take the regularized Net-brane $NB(\epsilon)$ of Fig. \ref{reNetbrane}. Then the intersection of $NB(\epsilon)$ and each edge $E_m$ behaves like a standard intersection of two surfaces, allowing us to incorporate Hayward terms as done in AdS/BCFT. Finally, we sum over $m$ and take the limit as $\epsilon$ approaches zero at the end of our calculations.

There are two typical options for $t_{ij}$ that satisfy $D^it_{ij}=0$. The first option is $t_{ij}=0$. The second option, related to holographic renormalization \cite{deHaro:2000vlm}, is given by: 
 \begin{align}\label{sect2: tij}
t^{ij}=\frac{2}{\sqrt{|h|}} \frac{\delta I_{ct}}{\delta h_{ij}}=-2(d-1)h^{ij}+\frac{2}{d-2}( \mathcal{R}^{ij}-\frac{\mathcal{R}}{2} h^{ij})+...
\end{align}
where $I_{ct}$ denotes the counterterm on the AdS boundary
 \begin{align}\label{sect2: Ic}
I_{ct}=\int_{E} d^dy\sqrt{|h|} \Big[ -2(d-1)-\frac{\mathcal{R}}{d-2}+... \Big]
\end{align}
Since the stress tensor $t_{ij}$ (\ref{sect2: tij}) is derived from a covariant action (\ref{sect2: Ic}), it automatically obeys $D^it_{ij}=0$. One can also check this with the first few terms of (\ref{sect2: tij}). With the second choice (\ref{sect2: tij}), $T_{ij}=2K h_{ij}-2K_{ij}+t_{ij}$ is the renormalized stress tensor
for NCFTs on edge $E_m$, while it is an auxiliary quantity for our proof on the Net-brane. 
Note that the Net-brane is asymptotically AdS near the node \footnote{Similar to AdS/BCFT, we focus on the AdS brane instead of dS or flat brane in AdS/NCFT.}. 
As a result, all the intrinsic curvatures on the Net-brane are proportional to the metric, and all the covariant derivatives vanish when it approaches the boundary of the Net-brane $N=\partial NB$. Thus, we have generally
 \begin{align}\label{sect2: tij NB}
t_{NB\ ij}|_N\sim h_{NB\ ij}|_N. 
\end{align}

Let us revisit the main idea of the proof and provide more details for clarity. 
With a conserved stress tensor $T_{ij}=2K h_{ij}-2K_{ij}+t_{ij}$ on edges and the Net-brane, we can construct a local conserved current with a local Killing vector $J_i = T_{ij} \xi^j$, where $ \xi^j\sim \delta^j_a$ and $a$ represents the tangential directions at the node $N = \partial NB = \partial E_m$. It is important to note that the spatial translation invariance normal to the node is broken on both the edges and the Net-brane. Therefore, $ \xi^j \sim \delta^j_n$ is no longer a Killing vector.  

\begin{figure}[htbp]
  \centering
\includegraphics[width=.4\textwidth]{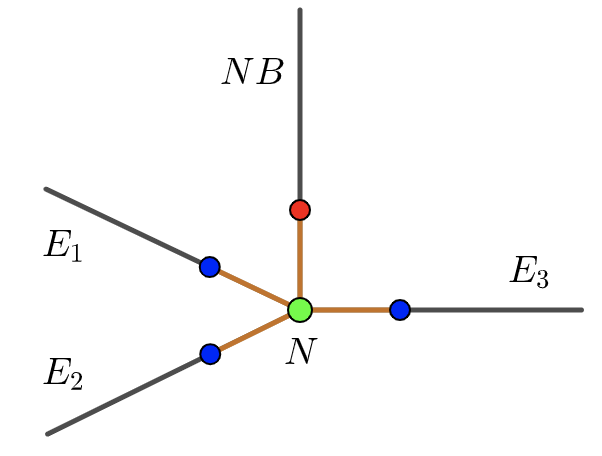}
\qquad 
\includegraphics[width=.5\textwidth]{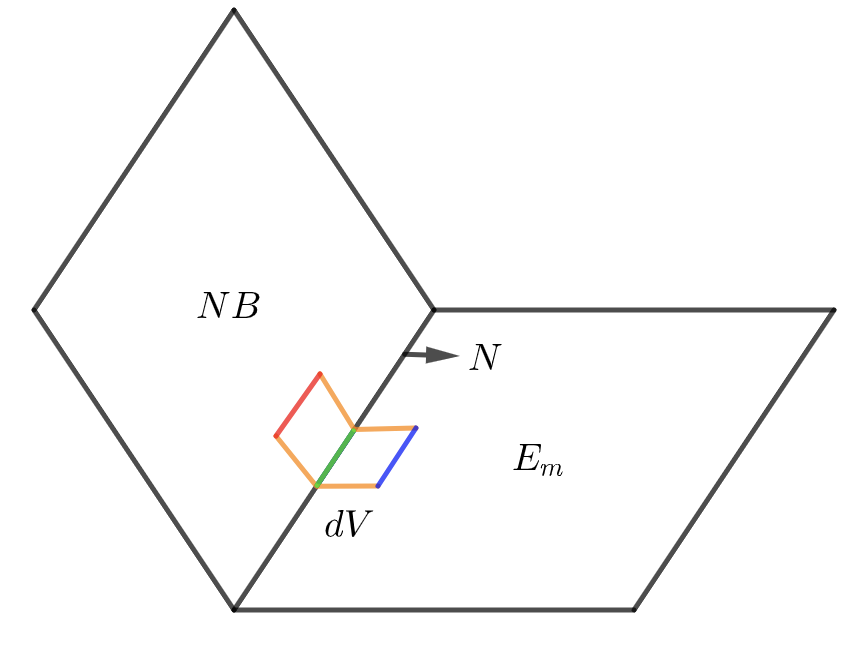}
 \caption{The region $V$, where we apply Gauss's law. This region $V$ is outlined by orange lines with red and blue endpoints in the left figure. The length of each orange line approaches zero ($ dl \to 0$), and the areas at the red, blue, and green points are all $dS$. 
In the right figure, we provide more details about $V$, which is bounded by the red, orange, and blue lines. The red, blue, and green points in the left figure correspond to the lines with the same colors in the right figure. }
  \label{proveNBC}
\end{figure}

Consider a small region $V$ that includes a piece of node $N$ with area $dS$ in edges $E_m$ and the Net-brane $NB$. This region $V$ is outlined by orange lines with red and blue endpoints, as shown in Fig. \ref{proveNBC} (left). The length of each orange line approaches zero ($ dl \to 0$), and the areas at the red, blue, and green points are all $dS$. In Fig. \ref{proveNBC} (right), we provide more details about $V$, which is bounded by the red, orange, and blue lines. The red, blue, and green points in Fig. \ref{proveNBC} (left) correspond to the lines of the same colors in Fig. \ref{proveNBC} (right). Without loss of generality, we focus on the Gauss normal coordinates. By applying Gauss's law in $ V $ and taking $dl \to 0$, we obtain:
\begin{align}\label{sect2: Gauss law}
0=\int_V  \nabla_i J^i dV= -\sum_{m}^p\overset{(m)}{J_E}_{\ n}|_{N} dS- \sum_{m}^p \overset{(m)}{J_{NB}}_{\ n} |_{N} dS,
\end{align}
where $J_E$ and $J_{NB}$ are the currents on edge $E$ and Net-brane $NB$. In our notation, n points from the edge/Net-brane to the node. For an arbitrary small area $dS$, the above equation yields 
\begin{align}\label{sect2: new proof 1}
 \sum_{m}^p \overset{(m)}{J_E}_{\ n}|_{N}=-\sum_{m}^p \overset{(m)}{J_{NB}}_{\ n}|_{N},
\end{align}
Since we have $J_n|_N = T_{ni} \xi^i|_N = T_{na}|_N$,  it follows that:
 \begin{align}\label{sect2: new proof 2}
 \sum_{m}^p \overset{(m)}{T_E}_{\ na}|_{N}= -\sum_{m}^p \overset{(m)}{T_{NB}}_{\ na}|_{N}.
\end{align}

Consider the first choice of conserved stress tensor given by $T_{ij}=2K h_{ij}-2K_{ij}$. The left-hand side (L.H.S) of (\ref{sect2: new proof 2}) is related to the NCFT stress tensors. The right hand side (R.H.S) of (\ref{sect2: new proof 2}) vanishes due to the junction condition (\ref{junction condition NCFT}) on the Net-brane, leading to the expression:
\begin{align}\label{sect2: physical argument 1}
\sum_m^p (2\overset{(m)}{K_E}\ h_{E\ na}-2\overset{(m)}{K_E}_{\ na})|_N=\sum_m^p (2\overset{(m)}{K_{NB}}_{\ na}-2\overset{(m)}{K_{NB}}\ h_{NB\ na})|_{N}=-2T h_{NB\ na}|_N=0.
\end{align}
Here, $i=(n,a)$ represents coordinates on the edge/Net-brane, with $n$ being the normal direction and $a$ the tangential direction to the node. Due to the orthogonality between $n$ and $a$, it follows that $h_{NB\ na}|_{N}=0$. Note that the junction condition (\ref{junction condition NCFT}) requires the sum over $m$, thereby making the sum in (\ref{sect2: physical argument 1}) essential. (\ref{sect2: physical argument 1}) implies that the total NCFT energy/tangential momentum fluxes into the node are zero. 

Next, consider the second choice  $T_{ij}=2K h_{ij}-2K_{ij}+t_{ij}$ with $t_{ij}$ given by (\ref{sect2: tij}). In this case, the L.H.S of (\ref{sect2: new proof 2}) represents the renormalized stress tensor of NCFT up to a factor: $T^{\text{CFT}}_{\ ij}=T_{ij}/\epsilon^{d}$, where $\epsilon$ is the cut-off AdS boundary $z=\epsilon$. The R.H.S of (\ref{sect2: new proof 2}) vanishes because of the junction condition (\ref{junction condition NCFT}) and the asymptotically AdS condition (\ref{sect2: tij NB}) on the Net-brane. Thus, we obtain
\begin{align}\label{sect2: physical argument 2}
\sum_m^p \overset{(m)}{T}{}^{\text{CFT}}_{\ na}|_N=\frac{1}{\epsilon^{d}}\sum_m^p (2\overset{(m)}{K_{NB}}_{\ na}-2\overset{(m)}{K_{NB}}\ h_{NB\ na}-t_{NB\ na})|_{N}\sim \frac{h_{NB\ na}}{\epsilon^{d}}|_N=0.
\end{align}
Note that $t_{NB\ ij}$ depends solely on the intrinsic curvatures and their covariant derivatives on the Net-brane, and is therefore independent of the branch index $m$. Note also that $h_{NB\ na}=0$ holds in all orders of $\epsilon$. This is true at least in Gauss normal coordinates. 

We have outlined the main idea of deriving the conservation law at a network node from the junction condition on the Net-brane. To establish a rigorous proof, we still need to tackle two problems. First, in general spacetimes, there are no global Killing vectors $\xi_i$, which means we do not have a globally conserved current $J_i = T_{ij}\xi^j$. This raises the question: does the key formula (\ref{sect2: new proof 2}) still hold if we only have a local Killing vector?  Second, we need to express $(2K_E h_{E\ na}-2K_{E\ na})$ of (\ref{sect2: physical argument 1}) in terms of the CFT stress tensor for the first choice. While related, these two expressions are generally different.

Now, we present a rigorous proof that overcomes the above two difficulties. As illustrated in Fig. \ref{proveNBC}, we select an infinitesimal region $V=dl dS$ that includes small portions of the Net-brane $NB$, the node $N$, and network edges $E_m$. This region is outlined by orange lines in Fig. \ref{proveNBC} (left), and is bounded by the red, orange, and blue lines in Fig. \ref{proveNBC} (right). We choose Gauss normal coordinates $y^i = (l, y^a)$ on both the edges and Net-brane, where $l$ is the direction normal to the node at $l=0$. Here $dl \to 0$ is the length of each orange line of \ref{proveNBC} (left), and the area $dS=d^{d-1}y$  remains the same for the red, green, and blue lines of Fig. \ref{proveNBC} (right). As discussed in Appendix \ref{app for Killing vectors}, we have a local Killing vector $\xi^i=\delta^i_a+O(y)$ that satisfies $D_{(i}\xi_{j)}=O(y)$ in Gauss normal coordinates. Based on this setup, we derive an approximate conserved current given by $J^i = T^{ij} \xi_j$. 

Integrating over the infinitesimal region $V$ gives us
\begin{align}\label{sect2: Gauss1}
&\int_{V} dV  D_i \Big(T^{ij}\xi_j\Big)=\int_{V} dV   T^{ij} D_{(i } \xi_{j)} \nonumber\\
& =\sum_{m}^p \int_{V\cap E_m} dV   \overset{(m)}{T}{}_E^{ij} D_{(i } \xi_{j)}+\sum_{m}^p \int_{V\cap NB} dV   \overset{(m)}{T}{}_{NB}^{ij} D_{(i } \xi_{j)} \nonumber\\
&\sim O(y) dV\sim O(y^{d+1})\to 0,
\end{align}
where we’ve utilized the fact that $D_{(i} \xi_{j)} \sim O(y)$ and neglected higher-order terms $O(y) dV=O(y)d^dy=O(y^{d+1})$. Here $dV=\sqrt{|h|}d^dy=O(y^d)$ with $\sqrt{|h|}\sim 1$ in Gauss normal coordinates. Applying Gauss's law, we find:
\begin{align}\label{sect2: Gauss2}
\int_{V} dV D_i \Big(T^{ij}\xi_j\Big)=-\sum_{m}^p\int_{S_{E_{m}}} dS   n^i \xi^j\overset{(m)}{T}_{E\ ij}-\sum_{m}^p\int_{S_{NB}} dS  n^i \xi^j\overset{(m)}{T}_{NB\ ij},
\end{align}
where $S_{NB}$ and $S_{E_m}$ represent the boundaries on the Net-brane and network edges, corresponding to the red and blue lines in Fig. \ref{proveNBC} (right), respectively. We ignore the lateral surfaces (orange line of Fig. \ref{proveNBC} (right)) as their contribution becomes negligible when $dl$ tends to zero. Note that the L.H.S of (\ref{sect2: Gauss2}) is $O(y^{d+1})$, while the R.H.S of (\ref{sect2: Gauss2}) is $O(y^{d-1})$ because $dS\sim O(y^{d-1})$ and $n^i=\delta^i_n+O(y)$, $\xi^i=\delta^i_a+O(y)$. Therefore, the R.H.S of (\ref{sect2: Gauss2}) must vanish at the order $O(y^{d-1})$. In the limit $dl\to 0$, we have $dS |_{E_m}\to dS |_{NB}\to dS |_{N}$, which means the red, green, and blue lines of Fig. \ref{proveNBC} (right) have the same area. Canceling this small area and using $n^i=\delta^i_n+O(y)$, $\xi^i=\delta^i_a+O(y)$, the R.H.S of (\ref{sect2: Gauss2}) results in the key equation (\ref{sect2: new proof 2}), which still holds at the node even with just local Killing vectors. 

Then, following discussions around (\ref{sect2: physical argument 2}), we prove that the junction condition on the Net-brane leads to the conservation on the network node, for the second choice of conserved stress tensor. Recall that we have used Gauss normal coordinates to argue the vanishing of the last term of (\ref{sect2: physical argument 2}), i.e., $ h_{NB\ na}/\epsilon^{d}|_N=0$. It is consistent with the above solid derivations of (\ref{sect2: Gauss1},\ref{sect2: Gauss2},\ref{sect2: new proof 2}), which focuses on the Gauss normal coordinates. 

We have now addressed the first problem concerning Killing vectors. Next, let us consider the second problem regarding the relationship between the extrinsic curvature \( K_E \) and the NCFT stress tensors for our first choice of conserved stress tensor. Note that (\ref{sect2: physical argument 1}) holds exactly since (\ref{sect2: new proof 2}) does. From (\ref{sect2: physical argument 1}) and $h_{E\ na}=0$, we get
\begin{align}\label{sect2: Gauss4}
\sum_{m}^p\overset{(m)}{K}_{E\ la} |_{N}=0,
\end{align}
where $l=-n$ denotes the normal direction.
Recall that we are working under the assumptions of Gauss normal coordinates. In the Fefferman-Graham (FG) gauge, the induced metric on the edges $ E_m$ (the AdS boundary at $z = \epsilon$) is given by
\begin{align}\label{sect2: FG gauge}
ds^2=\frac{(\eta_{ij}+ \frac{\epsilon^d}{d} T^{\text{CFT}}_{ij}+...)dx^i dx^j}{\epsilon^2},
\end{align}
where $x^i=(x, x^a)$ with the node at $x=0$, $T^{\text{CFT}}_{ij}$ represents the stress tensors of CFTs and $...$ denotes higher-order terms in $\epsilon$. For simplicity, we ignore the index $m$. 
Note that we focus on NCFTs in flat space with flat nodes. It means the CFT stress tensor \( T^{\text{CFT}}_{ij} \) remains finite at the node. In cases with a curved node, where the extrinsic curvatures have non-zero traceless parts \( \bar{k}_{ab} \), \( T^{\text{CFT}}_{ab} \sim \bar{k}_{ab}/x^{d-1} \) may become divergent \cite{Miao:2017aba}. However, this situation falls outside the scope of our paper. Given that we are working with a flat node where \( \bar{k}_{ab} = 0 \), we can safely utilize Gauss normal coordinates for both the edges and the Net-brane.
For the metric given above, we find that $ n^i \sim-\epsilon \delta^i_x$, and the extrinsic curvatures (from the bulk to the network edges) are 
\begin{align}\label{sect2: Kij1}
\hat{K}_{ij}\sim \frac{\eta _{ij}-\frac{(d-2)}{2 d} \epsilon ^d T^{\text{CFT}}_{ij}}{\epsilon ^2}. 
\end{align}
It is important to emphasize that $x^i = (x, x^a)$ are not the Gauss normal coordinates for non-zero $T^{\text{CFT}}_{xi}$. Transforming to the Gauss normal coordinates $y^i = (l, y^a)$, we have
\begin{align}\label{sect2: GN coordinates}
dl=(1+\frac{\epsilon^d}{2d}T^{\text{CFT}}_{xx} )\frac{dx}{\epsilon}+\frac{\epsilon^d}{d}T^{\text{CFT}}_{xa} \frac{dx^a}{\epsilon}+ O(\epsilon^{d}), \ dy^a=\frac{dx^a}{\epsilon}+O(\epsilon^{d-1}).
\end{align}
From this, we can derive the normal-tangential components of the extrinsic curvatures
\begin{align}\label{sect2: Kij}
K_{l a}=\frac{\partial x^i}{\partial l} \frac{\partial x^j}{\partial y^a} \hat{K}_{ij}\sim -\frac{\epsilon^{d}}{2}  T^{\text{CFT}}_{xa}. 
\end{align}
Substituting the expression for $K_{la}$ into (\ref{sect2: Gauss4}), we subsequently demonstrate that the junction condition on the Net-brane leads to the energy/tangential momentum conservation law on the network node locally
\begin{align}\label{sect2: proof}
\sum_{m}\overset{(m)}{T}{}^{\text{CFT}}_{na} |_{N}=0.
\end{align}
The following subsection will verify this conservation law in the perturbative Poincaré AdS. 

Under the $Z_p$ symmetry, AdS/NCFT with $p$ branches linked by one Net-brane simplifies to AdS/BCFT with an EOW brane. Correspondingly, the junction condition (\ref{junction condition NCFT}) on the Net-brane reduces to NBC (\ref{NBC BCFT}) on the EOW brane. Thus, the above discussions confirm that the AdS/BCFT with NBC (\ref{NBC BCFT}) is dual to the BCFT with a reflective boundary condition (\ref{BCFT BC}). On the other hand, since the Dirichlet boundary condition (DBC) fixed the induced metric rather than the extrinsic curvatures on the EOW brane, i.e., $K_{ij}-K h_{ij}\ne -T h_{ij}$, the key equations (\ref{sect2: physical argument 1},\ref{sect2: physical argument 2}) in the proof no longer hold. Consequently, the AdS/BCFT with DBC \cite{Miao:2018qkc} corresponds to the BCFT in an open system with a transparent boundary condition $J_n, T_{na}|_{\text{bdy}}\ne 0$ \footnote{We note that the AdS/BCFT with DBC fixes the tangential components of currents on the boundary, rather than the normal components. It is beyond the primary purpose of this paper and will be left for discussion in other papers.}. We will explore this interpretation and its relevance to open quantum systems in future articles. In section 2.3, we will show that AdS/BCFT with DBC contributes to the transparent modes of AdS/NCFT.

\subsection{Typical solutions}

This subsection studies some typical solutions for the holographic networks, starting with the simplest example with $p$ edges $E_m$ linked by one node $N$. The bulk metric in branch $B_m$ is given by 
\begin{align}\label{sect2: AdS metric}
{\text{Poincar\'e AdS}}:\ ds^2=\frac{dz^2+d\overset{(m)}{x}{}^2+\eta_{ab} dy^ady^b}{z^2},
\end{align}
where the network edge is at $z=0$, the node is at $\overset{(m)}{x}=z=0$ and the Net-brane is located at 
\begin{align}\label{sect2: AdS Nbrane}
\text{Net-brane}:\ \overset{(m)}{x}=-\sinh(\rho)z.
\end{align}
Note that we focus on the case where all bulk branches \(B_m\) share the same AdS radius, \(l_m=1\). More general cases will be discussed in future work.
Then the junction condition (\ref{junction condition NCFT}) on the Net-brane determines the brane tension
\begin{align}\label{sect2: AdS brane tension}
T=p(d-1)\tanh\rho. 
\end{align}
This Poincar\'e AdS is dual to the vacuum state of NCFTs on the simplest network. In the context of AdS/BCFT, the tension can be viewed as the gravitational counterpart to the boundary conditions in BCFTs, influencing the boundary central charges in a similar manner. Inspired by this, we anticipate that the tension of the Net-brane is related to the junction condition parameters for NCFTs in AdS/NCFT.

Let us discuss general networks that consist of at least two nodes or one loop. The Poincaré AdS (\ref{sect2: AdS metric}) is still a gravity dual for these networks, but does not represent the vacuum state \footnote{One should choose the brane tensions carefully to avoid the intersections of Net-branes. We focus on non-singular Net-branes in this paper.}. That is because Poincaré AdS has a holographic stress tensor of zero, $ \langle T_{ij}\rangle = 0 $, which cannot correspond to the vacuum of networks that have a non-trivial Casimir effect (where $ \langle T_{ij}\rangle \ne 0 $). To illustrate this, imagine a wave traveling along an edge between two nodes. When the wave reaches a node, some of it is transmitted, while some is reflected. It makes the node act like a partially transparent boundary, and the edge behaves like a translucent strip. Consequently, the Casimir effect here is weaker than that of a solid strip, yet it is still non-zero. This reasoning also applies to the network with loops. Please refer to \cite{Zhao:2025npv} for discussing the Casimir effect on networks. We expect that the gravity dual for the vacuum of general networks will involve the glue of Poincaré AdS for the external edges and AdS solitons for the internal edges. However, we will leave these discussions for future work and concentrate on simpler solutions in this paper.

Let us go on to discuss the gravity duals of thermal states of NCFTs. There are two typical solutions. 
The first one is the black string, which is dual to NCFT in a black hole background. 
For the simplest networks, the bulk metric in each branch $B_m$ reads 
\begin{align}\label{sect2: black string}
{\text{Black string}}:\ ds^2=d\overset{(m)}{r}{}^2+\cosh^2(\overset{(m)}{r}) \frac{\frac{dw^2}{h(w)}-h(w)dt^2+\delta_{ab} dy^a dy^b}{w^2},
\end{align}
where $h(w)=1-w^{d-1}/w_h^{d-1}$, the network edge is located at $\overset{(m)}{r}=-\infty$, the node is at $\overset{(m)}{r}=-\infty, w=0$, and the Net-brane with brane tension (\ref{sect2: AdS brane tension}) is at
\begin{align}\label{sect2: string Nbrane}
\text{Net-brane}:\ \overset{(m)}{r}=\rho.
\end{align}
The second one is the black hole, which is dual to the thermal state of NCFTs in flat space,
\begin{align}\label{sect2: black hole}
{\text{Black hole}}:\ ds^2=\frac{\frac{dz^2}{f(z)}-f(z) dt^2+d\overset{(m)}{x}{}^2+\delta_{ab} dy^a dy^b}{z^2}.
\end{align}
Here $f(z)=1-z^d/z_h^d$, the network edge is located at $z=0$, the node is at $z=x=0$, and the Net-brane is at
\begin{align}\label{sect2: string Nbrane}
\text{Net-brane}:\ \overset{(m)}{x}=0.
\end{align}
For the AdS black hole (\ref{sect2: black hole}), the junction condition (\ref{junction condition NCFT}) is obeyed only if the brane tension is zero, $T=0$. It is the main drawback compared to the black string (\ref{sect2: black string}). The advantage is that, like the Poincaré AdS, we can glue the AdS black hole solutions (\ref{sect2: black hole}) to get the gravity duals for the general networks. See the example in Fig. \ref{holo network} (right). 

To end this section, we will examine the perturbations of the AdS Poincaré metric and confirm that the junction condition (\ref{junction condition NCFT}) in bulk leads to the energy conservation law (\ref{NCFT BC}) at the node. For our analysis, we take the following ansatz for a perturbative metric
\begin{align}\label{sect2: AdS metric pert}
ds^2=\frac{dz^2-dt^2+d\overset{(m)}{x}{}^2+\epsilon \frac{2}{d} f_{m}(z,\overset{(m)}{x}) dt d\overset{(m)}{x}+\delta_{ab} dy^ady^b}{z^2},
\end{align}
where $\epsilon$ denotes the order of the perturbation, and 
$f_{m}(z,\overset{(m)}{x})$ represents the linear perturbation in the branch $B_m$. 
By solving Einstein's equations in linear order $O(\epsilon)$, we get
\begin{align}\label{sect2: ftx}
f_{m}(z,\overset{(m)}{x})=X_m(\overset{(m)}{x})+c_m z^d,
\end{align}
where $X_m(\overset{(m)}{x})$ is an arbitrary function. The junction condition (\ref{junction condition NCFT}) on the Net-brane (\ref{sect2: AdS Nbrane}) yields one independent equation
\begin{align}\label{sect2: ftx EOM1}
\cosh (\rho ) \sum_m c_m =0.
\end{align}
The continuity condition (\ref{continuity condition NCFT}) gives us
\begin{align}\label{sect2: ftx EOM2}
\sinh(\rho)\Big( c_m z^d+X_m(-z \sinh\rho )\Big)= c(z)
\end{align}
with $ c(z)$ a function independent of $c_m$. This continuity condition is satisfied automatically in the tensionless case $\rho=0$. In the tensive case $\rho\ne 0$, we solve $X_m(x)=b(x)-c_m x^d/(-\sinh(\rho))^d$ with $b(x)$ a general function of $x$, which is identical in every branch. 
Besides, we request $b(0)=0$ for $h_{xt}^{\text{CFT}}|_N=0$ on the node in every edges $E_{m}$.
From (\ref{sect2: FG gauge},\ref{sect2: AdS metric pert},\ref{sect2: ftx}), we derive the holographic stress tensor for NCFT
\begin{align}\label{sect2: holo Txt}
\overset{(m)}{T}{}^{\text{CFT}}_{xt}= \epsilon c_m.
\end{align}
Then the junction condition (\ref{sect2: ftx EOM1}) yields the conservation law for energy flux at the node
\begin{align}\label{sect2: conservation Txt}
\sum_m \overset{(m)}{T}{}^{\text{CFT}}_{xt}|_N= \epsilon \sum_m c_m=0.
\end{align}

Some comments are in order. First, as a perturbed metric, $f_{m}(z,\overset{(m)}{x})$ (\ref{sect2: ftx}) should remain small. Hence, the solution (\ref{sect2: AdS metric pert}) is invalid in the deep IR as $z \to \infty$. Second, we focus on the linear metric perturbation of order $O(\epsilon)$ in (\ref{sect2: AdS metric pert}). In this linear order, there is no naked singularity, as indicated by:
\begin{align}\label{reply 3: no singularity}
R_{\mu\nu\rho\sigma}R^{\mu\nu\rho\sigma}=2d(d+1)+O(\epsilon^2). 
\end{align}
Third, it is intriguing to explore non-perturbative solutions without naked singularities to verify the conservation law at the node. However, this problem exceeds the scope of the current paper, and we leave it for future work.

\subsection{Spectrum of KK modes}

This subsection studies the gravitational KK modes on the Net-brane. We focus on the simplest network with $p$ edges linked by one node for simplicity. In each bulk branch $B_m$, we take the following ansatz of the perturbative metric 
 \begin{align}\label{sect2: perturbative metric}
ds^2=dr^2+\cosh^2 (r) \left( \bar{h}^{(0)}_{ij}(y) + \epsilon \overset{(m)}{H}(r) \bar{h}^{(1)}_{ij}(y)  \right)dy^i dy^j+O(\epsilon^2),
\end{align}
where we ignore the index $(m)$ for $\overset{(m)}{r}$ in this subsection. Here, $\bar{h}^{(0)}_{ij}$ is the AdS metric with a unit radius and $\bar{h}^{(1)}_{ij}$ denotes the perturbation obeying the transverse traceless gauge 
 \begin{eqnarray}\label{sect2: hij1gauge}
\bar{D}^i \bar{h}^{(1)}_{ij}=0,\ \ \  \bar{h}^{(0)ij}\bar{h}^{(1)}_{ij}=0,
\end{eqnarray}
where $\bar{D}_i$ is the covariant derivative with respect to $\bar{h}^{(0)}_{ij}$. Separating variables of the linearized  Einstein equations, we obtain 
\begin{eqnarray}\label{sect2: EOMMBCmassivehij}
&& \left(\bar{\Box}+2-M^2\right)\bar{h}^{(1)}_{ij}(y)=0,\\
&& \cosh^2(r) \overset{(m)}{H}{}''(r)+d \sinh (r) \cosh (r) \overset{(m)}{H}{}'(r) + M^2 \overset{(m)}{H}(r)=0, \label{sect2: EOMMBCmassiveH}
\end{eqnarray}
where $M$ denotes the mass of gravitons on the Net-brane. Solving (\ref{sect2: EOMMBCmassiveH}) and imposing Dirichlet boundary condition (DBC) on the AdS boundary $\overset{(m)}{H}(-\infty)=0$, we get 
\begin{equation}\label{sect2: Htwocase}
\overset{(m)}{H}(r)= \overset{(m)}{c}H(r)= \overset{(m)}{c} \begin{cases}
 \ \text{sech}^{\frac{d}{2}}(r)  \ P_{l_M}^{\frac{d}{2}}(-\tanh r),&\
\text{even $d$} ,\\
\  \text{sech}^{\frac{d}{2}}(r) \ Q_{l_M}^{\frac{d}{2}}(-\tanh r),&\
\text{odd $d$}.
\end{cases}
\end{equation}
where $\overset{(m)}{c}$ are constants, $P$ and $ Q$ denote the Legendre polynomials, and $l_M$ is given by
 \begin{eqnarray}\label{sect2: aibia1}
l_M=\frac{1}{2} \left(\sqrt{(d-1)^2+4  M^2}-1\right).
\end{eqnarray}

We assume the location of the Net-brane remains invariant under linear perturbation
 \begin{align}\label{sect2: perturbative brane}
{\text{Net-brane}}: r=\rho+O(\epsilon^2). 
\end{align} 
Applying the junction condition (\ref{junction condition NCFT}) and the continuity condition (\ref{continuity condition NCFT}) on the Net-brane, we obtain 
 \begin{align}\label{sect2: per brane BC1}
& \sum_{m=1}^p \overset{(m)}{H}{}'(\rho)=0 \ \to   \ \sum_{m=1}^p \overset{(m)}{c}H'(\rho)= 0,\\
& \overset{(i)}{H}(\rho)= \overset{(j)}{H}(\rho) \ \to\  \ \ \ \overset{(i)}{c} H(\rho)=\overset{(j)}{c} H(\rho). \label{sect2: per brane BC2}
\end{align} 
Solving these boundary conditions gives us one class of modes that satisfy Neumann boundary conditions (NBC)
 \begin{align}\label{sect2: per brane NBC}
{\text{NBC}}:\ H'(\rho)=0, \ \overset{(i)}{c}=\overset{(j)}{c},
\end{align} 
and $(p-1)$ classes of modes that satisfy Dirichlet boundary conditions (DBC)
 \begin{align}\label{sect2: per brane DBC}
{\text{DBC}}:\ H(\rho)=0, \ \sum_{m=1}^p \overset{(m)}{c}=0. 
\end{align} 
Interestingly, the spectrum of AdS/NCFT combines characteristics from both AdS/BCFT with NBC \cite{Takayanagi:2011zk} and DBC/CBC \cite{Miao:2018qkc, Chu:2021mvq}. Here, CBC refers to the conformal boundary condition, which yields a spectrum identical to that of DBC \cite{Chu:2021mvq}. For discussions on the spectra of AdS/BCFT with various boundary conditions, please refer to \cite{Chu:2021mvq}; we will not repeat those details here. It is important to note that the spectrum is positive, satisfying $M^2 > 0$, and there exists an approximate massless mode with $M^2 \to 0$ in the large tension limit as $\rho \to \infty$.  
The spectrum $M^2$ can be calculated from Eqs. (\ref{sect2: Htwocase},\ref{sect2: aibia1}, \ref{sect2: per brane NBC},\ref{sect2: per brane DBC}). 
For the reader’s convenience, we provide the first few values of $M^2$ for $d=4$ and $\rho = -0.5, 0, 0.5$ in Tables \ref{tab: NBC} and \ref{tab: DBC}. The spectrum includes one class of non-degenerate modes obeying NBC and $(p-1)$ classes of degenerate modes obeying DBC. For both NBC and DBC, the values of $M^2$ are positive and decrease with increasing brane tension $ \rho$. 
\begin{table}[htbp] 
    \centering  
    \caption{Spectrum $M^2$ for KK modes with NBC and $d=4$}  
    \begin{tabular}{|c|c|c|c|}  
        \hline  
        $\rho=-0.5$ & 10.37 & 39.42 & 85.07\\  
        \hline
        $\rho=0$ & 4 & 18 &40\\
        \hline
        $\rho=0.5$ & 1.59 & 9.81 &22.71\\
        \hline
    \end{tabular}
    \label{tab: NBC}  
\end{table}
\begin{table}[htbp] 
    \centering  
    \caption{Spectrum $M^2$ for KK modes with DBC and $d=4$}  
    \begin{tabular}{|c|c|c|c|}  
        \hline  
        $\rho=-0.5$ & 21.32 & 58.71 & 112.67\\ 
        \hline
        $\rho=0$ & 10 & 28 & 54 \\
        \hline
        $\rho=0.5$ & 5.90 & 16.45 & 31.69 \\
        \hline
    \end{tabular}
    \label{tab: DBC}  
\end{table}

As discussed in section 2.1,  the AdS/BCFT with NBC corresponds to the BCFT with a reflective boundary condition $J_n, T_{na}|_N=0$. It indicates that the modes are confined to one edge and cannot flow to the other edges, which we refer to as isolated modes. On the other hand, the AdS/BCFT with DBC/CBC corresponds to an open system characterized by the transparent boundary condition $J_n, T_{na}|_N\ne 0$. In this case, the modes can flow freely between different edges; we call these modes transparent modes.

To summarize, we have derived the junction condition (\ref{junction condition NCFT}) on the Net-brane and proved that it leads to the conservation law (\ref{NCFT BC}) on the network node. We discuss several typical solutions to AdS/NCFT and verify the node conservation law (\ref{NCFT BC}) in the perturbative AdS. We find that the spectrum of gravitational KK modes on the Net-brane combines those of AdS/BCFT with NBC and DBC/CBC, which correspond to the isolated and transparent modes, respectively.

\section{Correlation functions}\label{sect for correlation function}

In this section, we study the two-point functions of scalar operators and stress tensors for the NCFTs on the simplest network with $p$ edges linked by one node. We first discuss the general constraint on two-point correlators from conformal symmetries and  conservation laws. Then, we study examples of free theories and the AdS/NCFT with tensionless Net-branes. Interestingly, these two special theories have similar structures of two-point functions. 

\subsection{General forms}

Let us outline the main ideas for deriving the general form of two-point functions. First, we will derive the reduced conformal group of NCFTs, which preserves the location of the node. Next, we will use the reduced conformal group to impose constraints on the general form of the correlators. This method is analogous to that used in BCFTs \cite{McAvity:1993ue}. A key difference for NCFTs is that the correlators contain more uncertain functions, which are constrained by the conservation law at the node. 

Let us start with the symmetric group of NCFT in a flat $d-$dimensional Euclidean space. We denote the coordinates of the network as $\overset{(m)}{\mathbf{x}}:~\overset{(m)}{x}{}^{i}=(\overset{(m)}{x}, \overset{(m)}{y}{}^a)$, where $m = 1, \ldots, p$ indicates different edges and $a = 2, \ldots, d$ represents the transverse directions. The node is situated at $\overset{(m)}{x} = 0$. Similar to BCFTs \cite{McAvity:1993ue}, the symmetry group of NCFTs that keeps the node location $\overset{(m)}{x} = 0$ unchanged is reduced from the full conformal group $O(d+1,1)$ to $O(d,1)$. The transformations for the $(d-1)$-dimensional translations, $O(d-1)$ rotations, and scale changes are given by
 \begin{align}\label{sect3: conformal group I}
 \overset{(m)}{y}{}^a \to  \overset{(m)}{y}{}^a+\xi^a,\  \overset{(m)}{y}{}^a\to R^a_{\ b} \overset{(m)}{y}{}^b, \ \overset{(m)}{x}{}^{i}\to \lambda \overset{(m)}{x}{}^{i}.
\end{align} 
To maintain the distance $|\overset{(m)}{\mathbf{x}}-\overset{(n)}{\mathbf{x}}|$ (up to a scale factor) between points on different edges, the transformations $\xi^a$, $R^a_{\ b}$, and $\lambda$ must take the same values in all edges. The special conformal transformations leaving $\overset{(m)}{x} = 0$ invariant read
 \begin{align}\label{sect3: conformal group II}
 \overset{(m)}{x}{}^{i}\to \frac{ \overset{(m)}{x}{}^{i}+b^{i} \overset{(m)}{\mathbf{x}}{}^2}{\Omega(\overset{(m)}{\mathbf{x}})}, \ \ \Omega(\mathbf{x})=1+2\mathbf{b}\cdot \mathbf{x}+\mathbf{b}^2 \mathbf{x}^2,
\end{align} 
where $b^{i}=(0, b^a)$ denotes the components of $\mathbf{b}$. 
For the same reason, $b^{i}$ takes the same values in all the edges. 

It is straightforward to construct two invariants under the above conformal transformations
 \begin{align}\label{sect3: invariant I}
&v_{\text{I}}^2=\frac{(\overset{(m)}{y}_a-\overset{(m)}{y}{}'_a)^2+(\overset{(m)}{x}-\overset{(m)}{x}{}')^2}{(\overset{(m)}{y}_a-\overset{(m)}{y}{}'_a)^2+(\overset{(m)}{x}+\overset{(m)}{x}{}')^2},\\
&v_{\text{II}}^2=\frac{(\overset{(m)}{y}_a-\overset{(n)}{y}{}'_a)^2+(\overset{(m)}{x}+\overset{(n)}{x}{}')^2}{(\overset{(m)}{y}_a-\overset{(n)}{y}{}'_a)^2+(\overset{(m)}{x}-\overset{(n)}{x}{}')^2}, \label{sect3: invariant II}
\end{align} 
where $v_{\text{I}}^2$ and $v_{\text{II}}^2$ are invariants on the same and mixed edges, respectively.  Note that, since $\overset{(m)}{x}\ge 0$, we have $|\overset{(m)}{\mathbf{x}}-\overset{(m)}{\mathbf{x}}{}'|^2=(\overset{(m)}{y}_a-\overset{(m)}{y}{}'_a)^2+(\overset{(m)}{x}-\overset{(m)}{x}{}')^2$, while $|\overset{(m)}{\mathbf{x}}-\overset{(n)}{\mathbf{x}}{}'|^2=(\overset{(m)}{y}_a-\overset{(n)}{y}{}'_a)^2+(\overset{(m)}{x}+\overset{(n)}{x}{}')^2$ in our notations. The two-point functions of scalar operators take the following form 
\begin{equation}\label{sect3: Htwocase}
\langle O(\mathbf{x})O(\mathbf{x}') \rangle= \begin{cases}
\frac{F_{\text{I}}(v_{\text{I}})}{|\overset{(m)}{\mathbf{x}}-\overset{(m)}{\mathbf{x}}{}'|^{2\Delta}},&\
\text{same edge} ,\\
\frac{F_{\text{II}}(v_{\text{II}})}{|\overset{(m)}{\mathbf{x}}-\overset{(n)}{\mathbf{x}}{}'|^{2\Delta}},&\ \text{mixed edge},
\end{cases}
\end{equation}
where $\Delta$ denotes the conformal dimension. In general, $F_{\text{I}}(v)$ and $F_{\text{II}}(v)$ are two independent functions. Due to the similar restricted conformal group $O(d,1)$, the two-point correlators take similar forms for NCFTs and BCFTs. The main difference is that NCFTs have more independent functions than BCFTs in the correlation functions.

Let us go on to discuss the two-point functions of stress tensors. The restricted conformal group $O(d,1)$ determines on the same edge \cite{McAvity:1993ue,Herzog:2017xha}
\begin{eqnarray}\label{sect3: TTfromHHnew}
 \langle  T_{ij}({\bf{x}})T_{kl}({\bf{x'}})  \rangle &=&\frac{1}{s^{2d}} \big[ \delta(v) \delta_{ij}\delta_{kl}+\epsilon(v) (I_{ik}I_{jl}+I_{il}I_{jk})+(\beta(v)-\delta(v))(\hat{X}_{i}\hat{X}_{j}\delta_{kl}+\hat{X}'_{k}\hat{X}'_{l}\delta_{ij})\nonumber\\
&&\ \ \ \ \ \  -\left(\gamma(v)+\epsilon(v)\right)(\hat{X}_{i}\hat{X}'_{k} I_{jl}+\hat{X}_{j}\hat{X}'_{l}I_{ik}+\hat{X}_{i}\hat{X}'_{l}I_{jk} +\hat{X}_{j}\hat{X}'_{k}I_{il}  ) \nonumber\\
&&+\left(\alpha(v)-2\beta(v)+4\gamma(v)+\delta(v)+2\epsilon(v) \right) \hat{X}_{i}\hat{X}_{j}\hat{X}'_{k}\hat{X}'_{l} \big],
\end{eqnarray}
where 
\begin{eqnarray}\label{sect3: Osbornformula}
\begin{split}
&\mathbf{s}=\mathbf{x}-\mathbf{x}',\mathbf{\bar{s}}=\mathbf{x}-\mathbf{\bar{x}}',~s=|\mathbf{s}|,~\bar{s}=|\bar{\mathbf{s}}|\\
&v=\frac{s}{\bar{s}}=\sqrt{\frac{(x-x')^2+(y_a-y'_a)^2}{(x+x')^2+(y_a-y'_a)^2}}, \\
&I_{ij}=\delta_{ij}-2\frac{(x_i-x'_i)(x_j-x'_j)}{s^2}, \\
&\hat{X}_i=\frac{1}{s\bar{s}}\left(x^2-x'^2-(y_a-y'_a)^2, 2x (y_a-y'_a) \right) ,\\
&\hat{X}'_i=\frac{1}{s\bar{s}}\left(x'^2-x^2-(y_a-y'_a)^2, -2x' (y_a-y'_a) \right),
\end{split}
\end{eqnarray} 
and $\bar{\mathbf{x}}'$ denotes the mirror point of $\mathbf{x}'$ about the node.

For simplicity, we have ignored $\overset{(m)}{ }$ for coordinates and ${}_I$ for all functions on the same edge. On the mixed edges, the two-point functions of stress tensors still take the form (\ref{sect3: TTfromHHnew}), with $v$ replaced by $v_{\text{II}}$ defined in (\ref{sect3: invariant II}), and $\mathbf{x}'=(x', y'_a)$ of $I_{ij}, \hat{X}_i, \hat{X}'_i$ replaced with $\bar{\mathbf{x}}'=(-x', y'_a)$.
Due to the network's symmetry, the functions $\alpha_{\text{I}}(v_{\text{I}}), \beta_{\text{I}}(v_{\text{I}}),...$ on the same edge are independent of the edge index $m$. The same applies to the functions   $\alpha_{\text{II}}(v_{\text{II}}), \beta_{\text{II}}(v_{\text{II}}),...$ on mixed edges. For simplicity, we will omit the index $m$ in the following discussion. Generally, the functions on the same edge are different from those on mixed edges, meaning $\alpha_{\text{I}} \ne \alpha_{\text{II}}, ...$. See examples (\ref{sect3: scalar alpha 1}) and (\ref{sect3: scalar alpha 2}) in the upcoming subsection. 

To be consistent with the conservation law (\ref{NCFT BC}), we require
\begin{eqnarray}\label{sect3: TT conservation}
\sum_{m} \lim_{\overset{(m)}{x}\to 0}  \langle \overset{(m)}{ T}_{na}({\bf{x}})\overset{(q)}{ T}_{kl}({\bf{x'}}) \rangle =0, 
\end{eqnarray} 
which yields the constraints
\begin{eqnarray}\label{sect3: TT constraint}
\gamma_{\text{I}}(v_{\text{I}}=1)-(p-1)\gamma_{\text{II}}(v_{\text{II}}=1)=0,
\end{eqnarray} 
where $p$ denotes the number of edges. 
Recall that $\gamma_{\text{I}}(v_{\text{I}})$ and $\gamma_{\text{II}}(v_{\text{II}})$ represent functions defined on the same and mixed edges, respectively, and these functions are generally different. To clarify the edge index, we can rewrite the equation as follows: 
\begin{align}\label{sect3: TT constraint 2} 
\overset{(qq)}{\gamma}(v_{\text{I}}=1) - (p-1) \overset{(m \neq q)}{\gamma}(v_{\text{II}}=1) = 0. 
\end{align} 
For the sake of simplicity, we will omit the cumbersome indices and adopt the simplified notation shown in (\ref{sect3: TT constraint}) in the following text.

In the following subsections, we will calculate the two-point functions of stress tensors for free NCFTs and AdS/NCFT with tensionless Net-branes. We verify that they all obey the constraints (\ref{sect3: TT constraint}).

\subsection{Free theories}

This subsection studies the two-point functions of stress tensors for free theories, including scalars, vectors and Dirac fermions.

\subsubsection{Free scalar }

We start with the Euclidean action for a conformally coupled free scalar
\begin{align} \label{sect3: scalar action}
    I=\frac{1}{2}\sum_{m=1}^p\int_{E_{m}} d^{d}x\ \delta^{ij} \partial_{i}\overset{(m)}{\phi} \partial_{j}\overset{(m)}{\phi},
\end{align}
where we have ignored the curvature term in flat space. Taking variations of the action, we get the following boundary term on the node 
\begin{align} \label{sect3: dI scalar}
   \delta I|_N=\sum_{m=1}^p\int_N d^{d-1}y\  \partial_{n} \overset{(m)}{\phi}\delta \overset{(m)}{\phi}=0,
\end{align}
which yields the following junction condition (JC), if we assume a continuous scalar field on the node,
\begin{align} \label{sect3: JC1 scalar}
    \text{scalar JC: }\ \ \ \sum_{m=1}^p\partial_{n}\overset{(m)}{\phi}|_N=0, \ \ \ \overset{(m)}{ \phi}|_N=\overset{(n)}{\phi}|_N.
\end{align}
The above JC is consistent with the conservation of energy flow on the node \cite{Zhao:2025npv} 
\begin{align}\label{sect3: conservation Tna}
\sum_{m=1}^p \overset{(m)}{T}_{na}|_N=0,
\end{align}
where $n,a$ denote the normal and tangential directions, and the stress tensors read
\begin{align} \label{sect3: scalar Tij}
    T_{ij}=&\partial_{i}\phi\partial_{j}\phi-\frac{1}{4}\frac{1}{d-1}\left( (d-2)\partial_{i}\partial_{j}+\delta_{ij}\partial^{2} \right)\phi^{2}\nonumber\\
    =&\frac{1}{2(d-1)}\left( d\partial_{i}\phi\partial_{j}\phi-(d-2)\phi\partial_{i}\partial_{j}\phi-\delta_{ij}(\partial\phi)^{2} \right).
  \end{align}
It is important to note that the JC (\ref{sect3: JC1 scalar}) is sufficient for ensuring energy conservation (\ref{sect3: conservation Tna}) at the node, but it is not the only option; there are other valid junction conditions available \cite{Pang:2025flq}. It is similar to the case of BCFTs, where there is generally more than one choice of boundary conditions. For simplicity, we focus on the JC (\ref{sect3: JC1 scalar}) with continuous fields at the node, and leave discussions of other JCs for future work. Additionally, keep in mind that scalar operators may not necessarily be continuous at the node. In the case of the JC (\ref{sect3: JC1 scalar}), the scalar field operator $\phi$ is continuous. However, for the bulk scalar that is continuous on the Net-brane, the corresponding dual operator is not continuous at the node. See Appendix \ref{app for holo one-point function} for more details. This scenario is similar to gravity, where the induced metric is continuous on the Net-brane, and the dual NCFT stress tensors satisfy the junction condition $\sum_{m} \overset{(m)}{T}_{na}|_{\text{node}}=0$. It typically implies a discontinuous stress tensor at the node. 
  
We solve the two-point function of scalars obeying the above JC as
\begin{equation} \label{sect3: scalar Green function}
\langle \phi(\mathbf{x})\phi(\mathbf{x}') \rangle=\frac{\kappa}{d-2}\cdot \begin{cases}
\frac{1+c_{r}v_{\text{I}}^{d-2}}{|\overset{(m)}{\mathbf{x}}-\overset{(m)}{\mathbf{x}}{}'|^{d-2}},&\
\text{same edge} ,\\
\frac{ c_{t}}{|\overset{(m)}{\mathbf{x}}-\overset{(n)}{\mathbf{x}}{}'|^{d-2}},&\ \text{mixed edge},
\end{cases}
\end{equation}
where 
\begin{align} \label{sect3: cr ct}
\kappa=\frac{\Gamma \left(\frac{d}{2}\right)}{2\pi ^{d/2}}, \ \ c_{r}=-\frac{p-2}{p}, \ \ c_{t}=\frac{2}{p}. 
   \end{align}
 From (\ref{sect3: scalar Green function},\ref{sect3: scalar Tij}), we derive the two-point function of stress tensors (\ref{sect3: TTfromHHnew}) with 
\begin{align}  \label{sect3: scalar alpha 1}
    \alpha_{\text{I}}(v_{\text{I}})=&\kappa^{2}\left( 1+c_{r}^{2}v_{\text{I}}^{2d}+ \frac{c_{r}}{4}(d-2)d\frac{d+1}{d-1}v_{\text{I}}^{d-2}(1-v_{\text{I}}^2)^{2} \right),\\  \label{sect3: scalar gamma 1}
    \gamma_{\text{I}}(v_{\text{I}})=&-\frac{\kappa^{2}d}{2(d-1)}\left( 1-c_{r}^{2}v_{\text{I}}^{2d}+\frac{c_{r}}{2}(d-2)\frac{d+1}{d-1}v_{\text{I}}^{d-2}(1-v_{\text{I}}^4) \right),\\
    \epsilon_{\text{I}}(v_{\text{I}})=&\frac{\kappa^{2}d}{2(d-1)}\left( 1+c_{r}^{2}v_{\text{I}}^{2d} \right)+\frac{c_{r}}{4}\frac{\kappa^{2}d}{(d-1)^{2}}\left( (d-2)(v_{\text{I}}^{d-2}+v_{\text{I}}^{d+2})+2dv_{\text{I}}^{d} \right),\label{sect3: scalar epsilon 1}
\end{align}
for the same edge and
\begin{align} \label{sect3: scalar alpha 2}
    \alpha_{\text{II}}(v_{\text{II}})=c_{t}^{2}\kappa^{2},~\gamma_{\text{II}}(v_{\text{II}})=-\frac{c_{t}^{2}\kappa^{2}d}{2(d-1)},~\epsilon_{\text{II}}(v_{\text{II}})=\frac{c_{t}^{2}\kappa^{2}d}{2(d-1)},
\end{align}
for the mixed edge.
Other functions of (\ref{sect3: TTfromHHnew}) can be derived with tracelessness of stress tensor \cite{McAvity:1993ue} as
\begin{align} \label{sect3: tracelessness}
    &\alpha+(d-1)\beta=0,~\beta+(d-1)\delta+2\epsilon=0.
\end{align}
We rewrite the two-point function of stress tensors into more familiar form
\begin{align} \label{sect3: scalar TT 1}
    \langle \overset{(m)}{ T}_{ij}({\mathbf{x}})\overset{(m)}{ T}_{kl}({\mathbf{x}'})\rangle=&C_{T}^{\phi}\left( \frac{\mathcal{I}_{ij,kl}(\mathbf{s})}{s^{2d}}+c_{r}^{2}\frac{\bar{\mathcal{I}}_{ij,kl}(\bar{\mathbf{s}})}{\bar{s}^{2d}}+c_{r}\frac{\#_{ij,kl}}{(s\bar{s})^{d}} \right),\\
       \langle \overset{(m)}{ T}_{ij}({\mathbf{x}})\overset{(n)}{ T}_{kl}({\mathbf{x}'})\rangle=&C_{T}^{\phi}\left(c_{t}^{2}\frac{\bar{\mathcal{I}}_{ij,kl}(\bar{\mathbf{s}})}{\bar{s}^{2d}}\right),~C_{T}^{\phi}=\frac{\kappa^{2}d}{d-1}  \label{sect3: scalar TT 2},
\end{align}
where the first term of (\ref{sect3: scalar TT 1}) agrees with that of a CFT. We have 
\begin{align}
    \mathcal{I}_{ij,kl}(\mathbf{s})=&\frac{1}{2}\left(I_{ik}(\mathbf{s})I_{jl}(\mathbf{s})+I_{il}(\mathbf{s})I_{jk}(\mathbf{s})\right)-\frac{1}{d}\delta _{ij}\delta_{kl}\\
    \bar{\mathcal{I}}_{ij,kl}(\bar{\mathbf{s}})=&\mathcal{I}_{ij,kl}(\mathbf{s})|_{x'\to\bar{x}'},
\end{align}
and $\#_{ij,kl}/(s \bar{s})^d$ comes from the mix of the CFT term $s$ and its mirror term $\bar{s}$. It is worth mentioning that $\#_{na,kl}$
 vanish in the limit $x\to0$. In such limit, we have $1-c_{r}^{2}=(p-1)c_{t}^{2}$ and can verify that the constraint (\ref{sect3: TT constraint}) on the node. Recall that we have $c_{r}=(2-p)/p$, $c_{t}=2/p$ and  $\gamma(v)$ is given by (\ref{sect3: scalar gamma 1},\ref{sect3: scalar alpha 2}).

\subsubsection{Maxwell theory}

 The Euclidean action of $4$-dimensional Maxwell theory reads
 \begin{align}\label{sect3: Maxwell action}
     I=\frac{1}{4}\sum_{m=1}^p\int_{E_{m}}d^{4}x  \overset{(m)}{F}_{ij} \overset{(m)}{F}{}^{ij},
 \end{align}
 whose variation yields 
  \begin{align}\label{sect3: Maxwell dI}
     \delta I|_N=\sum_{m=1}^p\int_{N}d^{3}y \overset{(m)}{F}{}^{na}\delta \overset{(m)}{A}_a=0.
 \end{align}
 From the above action variations, we read off the JC
  \begin{align}\label{sect3: Maxwell JC}
     \text{vector JC: }\ \  \sum_{m=1}^p\overset{(m)}{F}{}^{na}|_N=0, \ \ \ \overset{(m)}{F}{}_{ab}|_N=\overset{(n)}{F}{}_{ab}|_N.
 \end{align}
 We remark that the second equation of (\ref{sect3: Maxwell JC}) is the gauge invariant expression of the continuity condition $\overset{(m)}{A}_a=\overset{(n)}{A}_a$. Given the stress tensor 
  \begin{align}\label{sect3: Maxwell Tij}
     T_{ij}=F_{ik}F_{j}^{~k}-\frac{1}{4}\delta_{ij}F_{kl}F^{kl}, 
 \end{align}
it is easy to see that the vector JC (\ref{sect3: Maxwell JC}) yields the conservation law (\ref{sect3: conservation Tna}) on the node.

In the Feynman gauge, the vector two-point function obeying JC (\ref{sect3: Maxwell JC}) becomes
 \begin{equation} \label{sect3: Maxwell A1A1}
\langle A_{n}(\mathbf{x})A_{n}(\mathbf{x}') \rangle=\frac{\kappa}{2}\cdot \begin{cases}
\frac{1-c_{r} v_{\text{I}}^{2}}{|\overset{(m)}{\mathbf{x}}-\overset{(m)}{\mathbf{x}}{}'|^{2}},&\
\text{same edge} ,\\
\frac{- c_{t}}{|\overset{(m)}{\mathbf{x}}-\overset{(n)}{\mathbf{x}}{}'|^{2}},&\ \text{mixed edge},
\end{cases}
\end{equation}
and 
 \begin{equation}\label{sect3: Maxwell AaAb}
\langle A_{a}(\mathbf{x})A_{b}(\mathbf{x}') \rangle=\frac{\kappa\delta_{ab}}{2}\cdot \begin{cases}
\frac{1+c_{r}v_{\text{I}}^{2}}{|\overset{(m)}{\mathbf{x}}-\overset{(m)}{\mathbf{x}}{}'|^{2}},&\
\text{same edge} ,\\
\frac{ c_{t}}{|\overset{(m)}{\mathbf{x}}-\overset{(n)}{\mathbf{x}}{}'|^{2}},&\ \text{mixed edge},
\end{cases}
\end{equation}
which yields the two-point function of $F_{ij}$
\begin{align}\label{sect3: Maxwell FF}
    \langle F_{ij}(\mathbf{x})F_{kl}(\mathbf{x}')\rangle =&\frac{1}{s^{4}}\left[ 2a(v)(I_{ik}I_{jl}-I_{il}I_{jk})\right.\nonumber\\
    &~~~~\left. +b(v)\left(\hat{X}_{i}\hat{X}_{k}'I_{jl}-\hat{X}_{j}\hat{X}_{k}'I_{il}-\hat{X}_{i}\hat{X}_{l}'I_{jk}+\hat{X}_{j}\hat{X}_{l}'I_{ik}\right) \right],
\end{align}
with
\begin{align}\label{sect3: Maxwell ab1}
    a_{\text{I}}(v_{\text{I}})=\kappa (1+c_{r}v_{\text{I}}^{4}),~b_{\text{I}}(v_{\text{I}})=-4\kappa c_{r}v_{\text{I}}^{4},
\end{align}
for the same edge and
\begin{align}\label{sect3: Maxwell ab2}
    a_{\text{II}}(v_{\text{II}})=\kappa c_{t},~b_{\text{II}}(v_{\text{II}})=0,
\end{align}
for the mixed edge. Recall that $\mathbf{x}'=(x', y'_a)$ of $I_{ik},~\hat{X}_{i},~\hat{X}_{k}'$ should be replaced with $\bar{\mathbf{x}}'=(-x', y'_a)$ in the mixed edge. From (\ref{sect3: Maxwell FF}) and (\ref{sect3: Maxwell Tij}), we obtain the two-point functions of stress tensors (\ref{sect3: TTfromHHnew}) for Maxwell theory with 
\begin{align}\label{sect3: Maxwell alpha I}
    \alpha_{\text{I}}(v_{\text{I}})=12 \kappa ^2 \left( 1+c_{r}^{2}v_{\text{I}}^{8}\right),~\gamma_{\text{I}}(v_{\text{I}})=-8 \kappa ^2 \left( 1-c_{r}^{2}v_{\text{I}}^8\right),~\epsilon_{\text{I}}(v_{\text{I}})=8 \kappa ^2 \left( 1+c_{r}^{2}v_{\text{I}}^{8}\right),
\end{align}
for the same edge and
\begin{align}\label{sect3: Maxwell alpha II}
    \alpha_{\text{II}}(v_{\text{II}})=12\kappa^{2}c_{t}^{2},~\gamma_{\text{II}}(v_{\text{II}})=-8\kappa^{2}c_{t}^{2},~\epsilon_{\text{II}}(v_{\text{II}})=8\kappa^{2}c_{t}^{2},
\end{align}
for the mixed edge. From the above equations together with (\ref{sect3: cr ct}), we verify that the consistent condition (\ref{sect3: TT constraint}) is obeyed. Finally, we rewrite the two-point functions of stress tensors into more familiar form
\begin{align} \label{sect3: Maxwell TT 1}
 & \langle\overset{(m)}{ T}_{ij}({\mathbf{x}})\overset{(m)}{ T}_{kl}({\mathbf{x}'})\rangle=C_{T}^{A}\left( \frac{\mathcal{I}_{ij,kl}(\mathbf{s})}{s^{8}}+c_{r}^{2}\frac{\bar{\mathcal{I}}_{ij,kl}(\bar{\mathbf{s}})}{\bar{s}^{8}} \right),\\
  &\langle \overset{(m)}{ T}_{ij}({\mathbf{x}})\overset{(n)}{ T}_{kl}({\mathbf{x}'}) \rangle=C_{T}^{A}\left(c_{t}^{2}\frac{\bar{\mathcal{I}}_{ij,kl}(\bar{\mathbf{s}})}{\bar{s}^{8}}\right),~C_{T}^{A}=16\kappa^{2}. \label{sect3: Maxwell TT 2}
\end{align}
Interestingly, unlike the scalar, there are no cross-terms in $\langle TT \rangle $ for a vector.

 \subsubsection{Dirac fermions}

The Euclidean action of the free Dirac field reads in flat space
\begin{align}\label{sect3: Dirac action}
    I=&\int d^{d}x \bar{\psi}(\gamma^{i}\overset{\leftrightarrow}{\nabla}_{i})\psi\\
    =&\frac{1}{2}\sum_{m}\int_{E_{m}}d^{d}x \left(\bar{\psi}\gamma^{i}\partial_{i}\psi-(\partial_{i}\bar{\psi})\gamma^{i}\psi\right),
\end{align}
where $\gamma^{i}$ satisfies the Clifford algebra in Euclidean signature
\begin{align}\label{sect3: Dirac gamma}
    \{ \gamma_{i},\gamma_{j} \}=2\delta_{ij}. 
\end{align}
Following \cite{McAvity:1993ue}, we define the projectors $\Pi_{-}$ and $\Pi_{+}$
\begin{align}\label{sect3: Dirac projector}
    \Pi_{\pm}=\frac{1}{2}\left( \mathbf{1}\pm U \right), 
\end{align}
where $U=\Pi_{+}-\Pi_{-}$ obeys 
\begin{align}\label{sect3: Dirac chi}
    U\gamma_{n}=-\gamma_{n}\bar{U},~U\gamma_{a}=\gamma_{a}\bar{U},~U ^{2}=\bar{U}^{2}=\mathbf{1}.
\end{align}

The current and stress tensor for Dirac fields read
\begin{align}\label{sect3: Dirac Ji}
&J^i=i\bar{\psi}\gamma^{i} \psi, \\
&T_{ij}=\bar{\psi}\gamma_{(i}\overset{\leftrightarrow}{\nabla}_{j)}\psi\nonumber\\
&~~~~=\frac{1}{2}\left( \partial_{(i}\bar{\psi}\gamma_{j)}\psi-\bar{\psi}\gamma_{(i}\partial_{j)}\psi \right) \label{sect3: Dirac Tij}.
\end{align}
To satisfy the conservation law $\sum_m \overset{(m)}{J}_n=\sum_m \overset{(m)}{T}_{na}=0$ on the node, we choose the following JC
\begin{align}\label{sect3: Dirac JC}
\sum_{m}\Pi_{-}\overset{(m)}{\psi}|_N=\sum_{m}\overset{(m)}{\bar{\psi}}\bar{\Pi}_{-}|_N=0,\ \ \ \Pi_{+}\overset{(m)}{\psi}|_N=\Pi_{+}\overset{(n)}{\psi}|_N,~\overset{(m)}{\bar{\psi}}\bar{\Pi}_{+}|_N=\overset{(n)}{\bar{\psi}}\bar{\Pi}_{+}|_N,
\end{align}
where $\bar{\Pi}_{\pm}=\frac{1}{2}\left(\mathbf{1}\pm\bar{U}\right)$. The two-point function of Dirac field satisfying the above JC is given by

\begin{equation}\label{sect3: Dirac two point}
\langle \psi(\mathbf{x})\bar{\psi}(\mathbf{x}') \rangle=\kappa\cdot 
\begin{cases}
\frac{\gamma\cdot (\overset{(m)}{\mathbf{x}}-\overset{(m)}{\mathbf{x}}{}')}{|\overset{(m)}{\mathbf{x}}-\overset{(m)}{\mathbf{x}}{}'|^{d}}+c_{r}\frac{\gamma\cdot (\overset{(m)}{\mathbf{x}}-\overset{(m)}{\bar{\mathbf{x}}}{}')}{|\overset{(m)}{\mathbf{x}}-\overset{(m)}{\bar{\mathbf{x}}}{}'|^{d}}\bar{U},&\ \text{same edge} ,\\
c_{t}\frac{ \gamma\cdot (\overset{(m)}{\mathbf{x}}-\overset{(n)}{\mathbf{x}}{}')}{|\overset{(m)}{\mathbf{x}}-\overset{(n)}{\mathbf{x}}{}'|^{d}}\bar{U},&\ \text{mixed edge},
\end{cases}
\end{equation}
where $(\overset{(m)}{\mathbf{x}}-\overset{(n)}{\mathbf{x}}{}')=(x+x',y_{a}-y_{a}')$ denotes the physical separation between $\overset{(m)}{\mathbf{x}}$ and $\overset{(n)}{\mathbf{x}}{}'$ for $m\neq n$. 
Then, we derive the two-point functions of stress tensors 
\begin{align}\label{sect3: Dirac TT 1}
    <\overset{(m)}{ T}_{ij}({\mathbf{x}})\overset{(m)}{ T}_{kl}({\mathbf{x}'})>=&C_{T}^{\psi}\left( \frac{\mathcal{I}_{ij,kl}(\mathbf{s})}{s^{2d}}+c_{r}^{2}\frac{\bar{\mathcal{I}}_{ij,kl}(\bar{\mathbf{s}})}{\bar{s}^{2d}} \right),\\
    <\overset{(m)}{ T}_{ij}({\mathbf{x}})\overset{(n)}{ T}_{kl}({\mathbf{x}'})>=&C_{T}^{\psi}\left(c_{t}^{2}\frac{\bar{\mathcal{I}}_{ij,kl}(\bar{\mathbf{s}})}{\bar{s}^{2d}}\right),~C_{T}^{\psi}=\frac{d\kappa^{2}\text{tr}(\mathbf{1})}{2}, \label{sect3: Dirac TT 2}
\end{align}
where $\text{tr}(\mathbf{1})=2^{[d/2]}$. In the expressions of (\ref{sect3: TTfromHHnew}), we have 
\begin{align}\label{sect3: Dirac TT I}
    &\alpha_{\text{I}}(v_{\text{I}})=\frac{d-1}{2}\kappa^{2}\text{tr}(\mathbf{1})\left(1+c_{r}^{2}v_{\text{I}}^{2d}\right),~\gamma_{\text{I}}(v_{\text{I}})=-\frac{d}{4}  \kappa ^2\text{tr}(\mathbf{1}) \left(1-c_{r}^{2}v_{\text{I}}^{2d}\right), \nonumber\\
   & \epsilon_{\text{I}}(v_{\text{I}})=\frac{d}{4}  \kappa ^2 \text{tr}(\mathbf{1})\left(1+c_{r}^{2}v_{\text{I}}^{2d}\right),
\end{align}
for the same edge and
\begin{align}\label{sect3: Dirac TT II}
    \alpha_{\text{II}}(v_{\text{II}})=\frac{(d-1)}{2}\kappa^{2}\text{tr}(\mathbf{1})c_{t}^{2},~\gamma_{\text{II}}(v_{\text{II}})=-\frac{d}{4}  \kappa ^2\text{tr}(\mathbf{1})c_{t}^{2} ,~\epsilon_{\text{II}}(v_{\text{II}})=\frac{d}{4}  \kappa ^2 \text{tr}(\mathbf{1})c_{t}^{2},
\end{align}
for the mixed edge.

\subsection{Holographic theories}

This subsection investigates the two-point function of stress tensors in AdS/NCFT with a tensionless Net-brane. Following \cite{Liu:1998bu}, we take the ansatz for the metric perturbations as
\begin{align}\label{sect3: AdSNCFT metric}
    ds^{2}=\frac{dz^{2}+dx^{2}+\delta_{ab}dy^{a}dy^{b}+H_{\mu\nu}dx^{\mu}dx^{\nu}}{z^{2}},
\end{align}
with the gauge
\begin{align}\label{sect3: AdSNCFT gauge}
    H_{zz}(z=0,\textbf{x})=H_{zi}(z=0,\textbf{x})=0. 
\end{align}
Imposing the JC (\ref{junction condition NCFT}, \ref{continuity condition NCFT}) on the tensionless Net-brane and the DBC
\begin{align}\label{sect3: AdSNCFT DBC}
    H_{ij}(z=0,\textbf{x})=\hat{H}_{ij}(\textbf{x}),
\end{align}
on the AdS boundary, we obtain the linear metric perturbations
\begin{align}\label{sect3: AdSNCFT metric solution}
    \overset{(m)}{H}_{\mu\nu}(z,\textbf{x})=\frac{C_{T}}{2d}&\Bigg{\{}\int_{E_{m}} d^{d}x' \left[ \frac{z^{d}}{S^{2d}}J_{\mu i}J_{\nu j}P_{ijkl}+c_{r}\frac{z^{d}}{\bar{S}^{2d}}\bar{J}_{\mu i}\bar{J}_{\nu j}P_{ijkl} \right]\hat{H}_{kl}(\textbf{x}')\nonumber\\
    &+\sum_{n\neq m} c_{t}\int_{E_{n}}d^{d}x'\frac{z^{d}}{\bar{S}^{2d}}\bar{J}_{\mu i}\bar{J}_{\nu j}P_{ijkl}\hat{H}_{kl}(\textbf{x}')\Bigg{\}},
\end{align}
where
\begin{align}\label{sect3: AdSNCFT metric solution 123}
    C_{T}=&\frac{2\Gamma [d+2]}{\pi^{d/2}\Gamma [d/2](d-1)},\nonumber\\
    S^{2}=&z^{2}+(x-x')^{2}+(y_{a}-y_{a}')^{2},\nonumber\\
    \bar{S}^{2}=&z^{2}+(x+x')^{2}+(y_{a}-y_{a}')^{2},\nonumber\\
    P_{ijkl}=&\frac{1}{2}\left( \delta_{ik}\delta_{jl}+\delta_{il}\delta_{jk} \right)-\frac{1}{d}\delta_{ij}\delta_{kl},\\
    J_{\mu \nu}=&\delta_{\mu\nu}-2\frac{(x_{\mu}-x_{\mu}')(x_{\nu}-x_{\nu}')}{S^{2}},\nonumber\\
    \bar{J}_{\mu\nu}=&J_{\mu\nu}-2X_{\mu}X_{\nu}',\nonumber
\end{align}
and
\begin{align}\label{sect3: AdSNCFT metric solution XbarX}
    X_{\mu}=&\frac{1}{S\bar{S}}\left( 2xz,x^{2}-x'^{2}-(y_{a}-y_{a}')^{2}-z^{2},2x(y_{a}-y_{a}') \right),\\
    \bar{X}_{\mu}=&\frac{1}{S\bar{S}}\left( -2x'z,x'^{2}-x^{2}-(y_{a}-y_{a}')^{2}-z^{2},-2x'(y_{a}-y_{a}') \right). \label{sect3: AdSNCFT metric solution barX}
\end{align}

According to \cite{Liu:1998bu}, the on-shell quadratic action for $H_{ij}$ is given by
\begin{align}\label{sect3: AdSNCFT quadratic action}
    I_{2}=&\sum_{m}\int_{E_{m}}d^{d}xz^{1-d}\left( \frac{1}{4} H_{ij}\partial_{z}H_{ij}-\frac{1}{2}H_{ij}\partial_{j}H_{zi} \right)\nonumber\\
    =&\frac{C_{T}}{8}\Bigg{\{}\sum_{m}\int_{E_{m}}d^{d}xd^{d}x'  \hat{H}_{ij}(\mathbf{x})\left[ \frac{\mathcal{I}_{ij,kl}}{s^{2d}}+c_{r}\frac{\bar{\mathcal{I}}_{ij,kl}}{\bar{s}^{2d}} \right]\hat{H}_{kl}(\mathbf{x}')\nonumber\\
    &~~~~~+\sum_{m}\sum_{n\neq m}\int_{E_{m}}d^{d}x\int_{E_{n}}d^{d}x'\hat{H}_{ij}(\mathbf{x})c_{t}\frac{\bar{\mathcal{I}}_{ij,kl}}{\bar{s}^{2d}} \hat{H}_{kl}(\mathbf{x}')\Bigg{\}},
\end{align}
where
\begin{align}\label{sect3: AdSNCFT Iijkl}
    \mathcal{I}_{ij,kl}=&\lim_{z\to0}\frac{1}{2}\left( J_{ik}J_{jl}+J_{il}J_{jk} \right)-\frac{1}{d}\delta_{ij}\delta_{kl},\\
    \bar{\mathcal{I}}_{ij,kl}=&\lim_{z\to0}\frac{1}{2}\left( \bar{J}_{ik}\bar{J}_{jl}+\bar{J}_{il}\bar{J}_{jk} \right)-\frac{1}{d}\delta_{ij}\delta_{kl}.
\end{align}
From (\ref{sect3: AdSNCFT quadratic action}), we obtain the holographic two-point function of stress tensor
\begin{align} \label{sect3: holo TT I}
    \langle \overset{(m)}{ T}_{ij}({\mathbf{x}})\overset{(m)}{ T}_{kl}({\mathbf{x}'}) \rangle =&C_{T}\left( \frac{\mathcal{I}_{ij,kl}(\mathbf{s})}{s^{2d}}+c_{r}\frac{\bar{\mathcal{I}}_{ij,kl}(\bar{\mathbf{s}})}{\bar{s}^{2d}} \right),\\
    \langle  \overset{(m)}{ T}_{ij}({\mathbf{x}})\overset{(n)}{ T}_{kl}({\mathbf{x}'})   \rangle =&C_{T}\left(c_{t}\frac{\bar{\mathcal{I}}_{ij,kl}(\bar{\mathbf{s}})}{\bar{s}^{2d}}\right)  \label{sect3: holo TT II}.
\end{align}
We remark that (\ref{sect3: holo TT I}) agrees with that of AdS/BCFT \cite{Miao:2018dvm} for $p=1$ and $c_r=(2-p)/p=1$, which is a test of our calculations. Finally, we rewrite the above equations into the form of (\ref{sect3: TTfromHHnew}) with
\begin{align} \label{sect3: holo TT alpha I}
    \alpha_{\text{I}}(v_{\text{I}})=\frac{(d-1)}{d}C_{T}(1+c_{r}v_{\text{I}}^{2d}),~\gamma_{\text{I}}(v_{\text{I}})=-\frac{C_{T}}{2} (1-c_{r}v_{\text{I}}^{2d}),~\epsilon_{\text{I}}(v_{\text{I}})=\frac{C_{T}}{2} (1+c_{r}v_{\text{I}}^{2d}),
\end{align}
for the same edge and
\begin{align} \label{sect3: holo TT alpha II}
    \alpha_{\text{II}}(v_{\text{II}})=\frac{(d-1)}{d}C_{T}c_{t},~\gamma_{\text{II}}(v_{\text{II}})=-\frac{C_{T}}{2}c_{t},~\epsilon_{\text{II}}(v_{\text{II}})=\frac{C_{T}}{2}c_{t},
\end{align}
for the mixed edge. We verify that they obey the constraints (\ref{sect3: TT constraint}) from energy conservation.

In summary, we have examined the general form of two-point functions for NCFTs. They share similar expressions with BCFTs but involve more independent functions. We also explored examples of free theories and the AdS/NCFT with tensionless branes.

Some comments are for the correlators of free NCFTs and AdS/NCFT. 

\begin{itemize}

 \item It is interesting to note that different NCFTs include the same elements $c_{r}=(2-p)/p$ and $c_{t}=2/p$ in the correlation functions (\ref{sect3: scalar TT 1},\ref{sect3: scalar TT 2},\ref{sect3: Maxwell TT 1},\ref{sect3: Maxwell TT 2},\ref{sect3: Dirac TT 1},\ref{sect3: Dirac TT 2},\ref{sect3: holo TT I},\ref{sect3: holo TT II}). We interpret $c_r$ and $c_t$ as the reflection and transmission coefficients, respectively.  A strong support for this interpretation is the equality
\begin{align}\label{sect3: probability conservation}
   c_r^2+(p-1)c_t^2=(\frac{2-p}{p})^2+(p-1)(\frac{2}{p})^2=1,
\end{align}
where $c_r^2$ represents reflectance, $c_t^2$ is the transmittance to one edge. For a node connecting $p$ edges, the total transmittance is $(p-1)c_t^2$.  

 \item It is noteworthy that  these reflection and transmission coefficients depend solely on the network structure, rather than the types of fields involved. It may result from similar junction conditions for various fields. Recall that we impose the universal junction condition, where the field is continuous, and the sum of normal derivatives of the field at the node is zero. For a wave function moving from edge $E_1$ to the node at $\overset{(m)}{x}=0$, we have
\begin{align}\label{sect3: cr ct meaning 1}
  \overset{(1)}{X}=1\ e^{-i k \overset{(1)}{x}-i \omega t}+c_r \ e^{i k \overset{(1)}{x}-i \omega t},
\end{align}
where the first term is the incident wave and the second term is the reflected wave. The transmitted wave to the edge $E_m$ is 
\begin{align}\label{sect3: cr ct meaning 2}
  \overset{(m)}{X}= c_t\  e^{i k \overset{(m)}{x}-i \omega t}, \ \ m=2, 3, ..., p,
\end{align}
where $k=\omega$ and $X$ labels general fields. Imposing the junction condition at the node gives:
\begin{align}\label{sect3: cr ct meaning 3}
  \overset{(m)}{X}=  \overset{(n)}{X}, \ \ \ \sum_{m}^p \partial_{ \overset{(m)}{x}}  \overset{(m)}{X}=0,
\end{align}
which leads to the derived coefficients $c_{r}=(2-p)/p$ and $c_{t}=2/p$. Since $X$ is the wave function, the squares $c_{r}^2$ and $c_{t}^2$ are the reflection and transmission probabilities. As a result, (\ref{sect3: probability conservation}) means the probability conservation. 

 \item Note that we focus on JC (\ref{sect3: cr ct meaning 3}) in this paper. For an ICFT, there are more general choices of JCs \cite{Bachas:2001vj,Meineri:2019ycm,Quella:2007hr,Delfino:1994np}. Take the $(1+1)$-dimensional free scalar as an example, the general JC reads
 \begin{align} \label{sect3: general JC ICFT}
    \text{general JC: }\left\{\begin{array}{cc}
         \partial_{t}\overset{(1)}{\phi}=\frac{1}{2}\left( (1\pm 1)\partial_{t}\overset{(2)}{\phi}+(1\mp 1)\partial_{x}\overset{(2)}{\phi} \right)\cot{\theta} , \\
         \partial_{x}\overset{(1)}{\phi}=-\frac{1}{2}\left((1\mp 1)\partial_{t}\overset{(2)}{\phi}+(1\pm 1) \partial_{x}\overset{(2)}{\phi}\right)\tan{\theta}.
    \end{array}\right.
\end{align}
By taking $+,-$ of $\pm,\mp$, it encompasses the JC (\ref{sect3: cr ct meaning 3})
as a specific case with $\theta=\frac{\pi}{4}$.
The transmittance for the general JC (\ref{sect3: general JC ICFT}) is given by \cite{Bachas:2001vj,Meineri:2019ycm,Quella:2007hr}
 \begin{align} \label{sect3: general transmittance}
 \mathcal{T}=(\sin2\theta)^2, 
\end{align}
 which aligns with our special case $\mathcal{T}=1$ with \(\theta = \frac{\pi}{4}\) for JC (\ref{sect3: cr ct meaning 3}) with $p=2$. These above discussions suggest that, beyond JC  (\ref{sect3: cr ct meaning 3}), there are more general junction conditions for NCFTs with $p>2$ as well. We intend to address these general junction conditions in future work.

\item Note that the dependencies of $c_r$ and $c_t$ in correlators vary for different fields due to the lesser conformal symmetry of NCFTs compared to CFTs. Therefore, we do not expect the two-point functions to have universal forms across all fields.

\item  The correlators of NCFTs rely on the reflection coefficient $c_{r}=(2-p)/p$ and transmission coefficient $c_{t}=2/p$, where $p$ is the number of edges. Setting $p=1$ yields the correlators for BCFTs, highlighting the distinction between a node ($p>1$) and a boundary ($p=1$). We have verified that our results are consistent with those of free BCFTs \cite{McAvity:1993ue, Herzog:2017xha} and AdS/BCFT with a tensionless brane \cite{Miao:2018dvm}. 

\item The one-point functions for certain operators can be determined from the two-point functions in our setup. For simplicity, we focus on the one-point function of scalar operators. We have for the scalar field 
\begin{align}\label{sect3: one point function scalar}
\langle \phi^2(\mathbf{x})\rangle|_{E_m}=\lim_{\mathbf{x}'\to \mathbf{x}} \langle \phi(\mathbf{x}) \phi(\mathbf{x}') \rangle|_{E_m}=\frac{\kappa}{d-2}\frac{c_r}{(2\overset{(m)}{x})^{d-2}},
\end{align}
for the $4d$ Maxwell theory
\begin{align}\label{sect3: one point function vector}
\langle F_{ij}(\mathbf{x})F^{ij}(\mathbf{x})\rangle|_{E_m}=\lim_{\mathbf{x}'\to \mathbf{x}}\langle F_{ij}(\mathbf{x})F^{ij}(\mathbf{x}')\rangle|_{E_m}=24\kappa\frac{ c_{r}}{(2\overset{(m)}{x})^{4}},
\end{align}
and for the Dirac fermions
\begin{align}\label{sect3: one point function fermion}
\langle \bar{\psi}(\mathbf{x})\psi(\mathbf{x})\rangle|_{E_m}=\lim_{\mathbf{x}'\to \mathbf{x}}\langle \bar{\psi}(\mathbf{x})\psi(\mathbf{x}')\rangle|_{E_m}=-\kappa \frac{c_{r}\text{tr}(U\gamma_{n})}{(2\overset{(m)}{x})^{d-1}}. 
\end{align} 
Similar to BCFTs, we have removed the contributions from free space without nodes, allowing us to find a finite one-point function. Interestingly, the above one-point function takes the universal form 
\begin{align}\label{sect3: scalar operator}
\langle O(\mathbf{x})\rangle|_{E_m}=\frac{a_m}{\overset{(m)}{x}{}^{\Delta}},
\end{align} 
where $\Delta$ is the operator's conformal dimension and $a_m$ is a constant. 

See Appendix \ref{app for holo one-point function} for the holographic discussion of the one-point function (\ref{sect3: scalar operator}).  
 \end{itemize}

\section{Holographic entanglement entropy}
\label{sect for HEE}

This section investigates holographic entanglement entropy (HEE) in AdS/NCFT. We propose that the RT surface connects solely to the Net-brane for a connected subsystem within the network. We derive the junction condition for Ryu-Takayanagi (RT) surfaces and confirm that our proposal is consistent with the strong subadditivity of entanglement entropy. Lastly, we explore several natural proposals for network entropy and analyze their behaviors under node RG flow.

\subsection{General theories}

\begin{figure}[!h]
    \centering
    \includegraphics[width=0.6\linewidth]{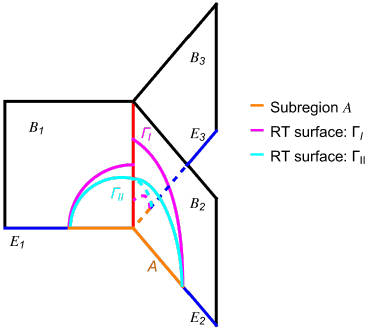}
    \caption{Two proposals of RT surfaces. The orange lines illustrate a typical choice for the subregion $A$. The magenta curves, \(\Gamma_{\text{I}}\), represent the RT surfaces from the first proposal, which is discontinuous at the Net-brane and intersects it orthogonally. In contrast, the cyan-blue curves $\Gamma_{\text{II}}$ correspond to the second proposal, where the RT surfaces are continuous across the Net-brane but not necessarily orthogonal to it.}
    \label{RT proposals}
\end{figure}

Let us analyze the simplest network consisting of three edges connected by a single node. There are two natural proposals regarding the entanglement entropy (HEE) of a connected subsystem that includes the node within this network. See Fig. \ref{RT proposals}. The first proposal, based on the AdS/BCFT framework, suggests that the RT surfaces are perpendicular to the Net-brane and may be disconnected. The second proposal, however, asserts that all RT surfaces must be interconnected through the same intersection on the Net-brane. While the first proposal leads to a smaller HEE, the second approach is ultimately the correct one for the following reasons.

\begin{itemize}
 \item Naturally, a connected subsystem of NCFT corresponds to a connected RT surface in bulk. It is essential to note that a network node differs from a boundary; information and energy can pass through a node and flow into different network edges. Consequently, the Net-brane dual to a node is a link rather than the end of spacetime. Therefore, physical continuity necessitates the presence of a connected RT surface in the bulk.
 
 \item AdS/ICFT (interface CFT) is a specific case of AdS/NCFT that involves only two edges and one node. When considering AdS/ICFT with a tensionless brane, it reduces to the standard AdS/CFT. In this context, the corresponding RT surface should be connected as in AdS/CFT. It indisputably rules out the disconnected proposal for HEE in AdS/NCFT.  
 
 \item The second proposal of HEE obeys the strong subadditivity of entanglement entropy. Let's consider two subsystems, $A_{a}$ (shown with orange lines) and $A_{b}$ (shown with green lines), along with their corresponding RT surfaces, $\Gamma_a$ and $\Gamma_b$, as depicted in Fig. \ref{SSA:1}. By combining these RT surfaces, we obtain the extremal surfaces $\tilde{\Gamma}_{\cup}$ and $\tilde{\Gamma}_{\cap}$ for the combined subsystems $A_{a}\cup A_{b}$ and the intersection $A_{a}\cap A_{b}$, illustrated in Fig. \ref{SSA:2}. Since RT surfaces are minimal, their areas are less than those of the extremal surfaces $\tilde{\Gamma}_{\cup}$ and $\tilde{\Gamma}_{\cap}$, which equals the total areas of $\Gamma_a$ and $\Gamma_b$. In this way, we prove subadditivity in AdS/NCFT
  \begin{align}\label{sect4: SSA}
     S(A_{a})+S(A_{b})=\frac{\text{Area}_{\tilde{\Gamma}_{\cup}}+\text{Area}_{\tilde{\Gamma}_{\cap}}}{4G_{N}}\geq S(A_{a}\cup A_{b})+S(A_{a}\cap A_{b}).
 \end{align}
 \begin{figure}[htbp]
  \centering
  \begin{subfigure}[b]{0.36\textwidth}
    \centering
    \includegraphics[width=\textwidth]{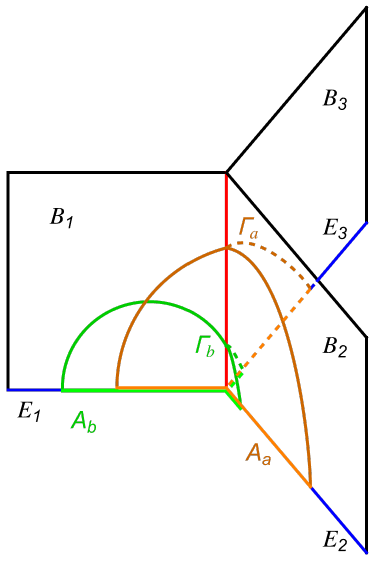}
    \caption{Subsystems and the RT surfaces}
    \label{SSA:1}
  \end{subfigure}
  \hspace{0.1\textwidth} 
  \begin{subfigure}[b]{0.36\textwidth}
    \centering
    \includegraphics[width=\textwidth]{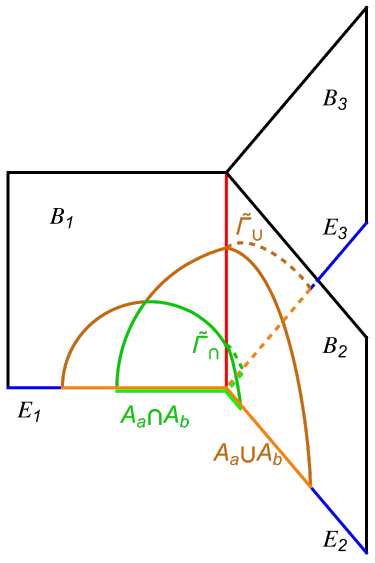}
    \caption{Recombination of RT surfaces}
    \label{SSA:2}
  \end{subfigure}
  \caption{Left: subsystems $A_{a}$ (orange lines) and $A_{b}$ (green lines) on edges and the corresponding RT surfaces $\Gamma_a$ (orange curves) and $\Gamma_b$ (green curves) in bulk.
    Right: subsystems $A_{a} \cup A_b$ (orange lines) and $A_{a} \cap A_b$ (green lines) and the corresponding extremal surfaces $\tilde{\Gamma}_{\cup}$ (orange curves) and $\tilde{\Gamma}_{\cap}$ (green curves), which combines the RT surfaces of $\Gamma_a$ and $\Gamma_b$ .} 
  \label{SSA:12}
\end{figure}
 
 \item Last but not least, only the second proposal aligns with the monotonicity of entanglement entropy for networks containing internal edges. Consider a finite subregion $A$ that includes at least two nodes and one internal edge within an infinite network. When we increase the length of the internal edge or add more internal edges, the entanglement between these edges and the complement $\bar{A}$ increases. Consequently, the entanglement entropy should also rise with the length or number of internal edges. It holds for the second proposal of HEE, which we will demonstrate below. In contrast, for the first proposal, the disconnected RT surface associated with the internal edge approaches infinity in the bulk, resulting in zero area. Therefore, the corresponding HEE is irrelevant to the length or number of internal edges, which is unphysical.
 
 \end{itemize}
 
Let us prove that the second proposal of HEE increases when we add a new internal edge to the connected subsystem $A$. Without loss of generality, we focus on the Poincar\'e AdS with tensionless Net-branes in bulk. We also assume that the subsystem $A$ is much smaller than its complement $\bar{A}$. We can simplify this problem by considering the opposite scenario: removing an internal edge. Label the network with $p$ internal edges and the corresponding HEE as $\text{Net}(p)$ and $S(p)$. When we remove one internal edge while keeping the RT surfaces of the other edges the same, the areas of the extremal surfaces $S_0(p-1)$ decrease. Thus, we have $S(p) > S_0(p-1)$. It is important to note that $S_0(p-1)$ is not the same as the HEE $S(p-1)$ for the network $\text{Net}(p-1)$, because the RT surfaces can change between $\text{Net}(p)$ and $\text{Net}(p-1)$. Since RT surfaces are minimal, we generally have $S_0(p-1) \ge S(p-1)$. It leads us to prove the inequality
\begin{align}\label{sect4: prove second entropy}
S(p)> S(p-1). 
\end{align} 

\begin{figure}
    \centering
    \includegraphics[width=0.7\linewidth]{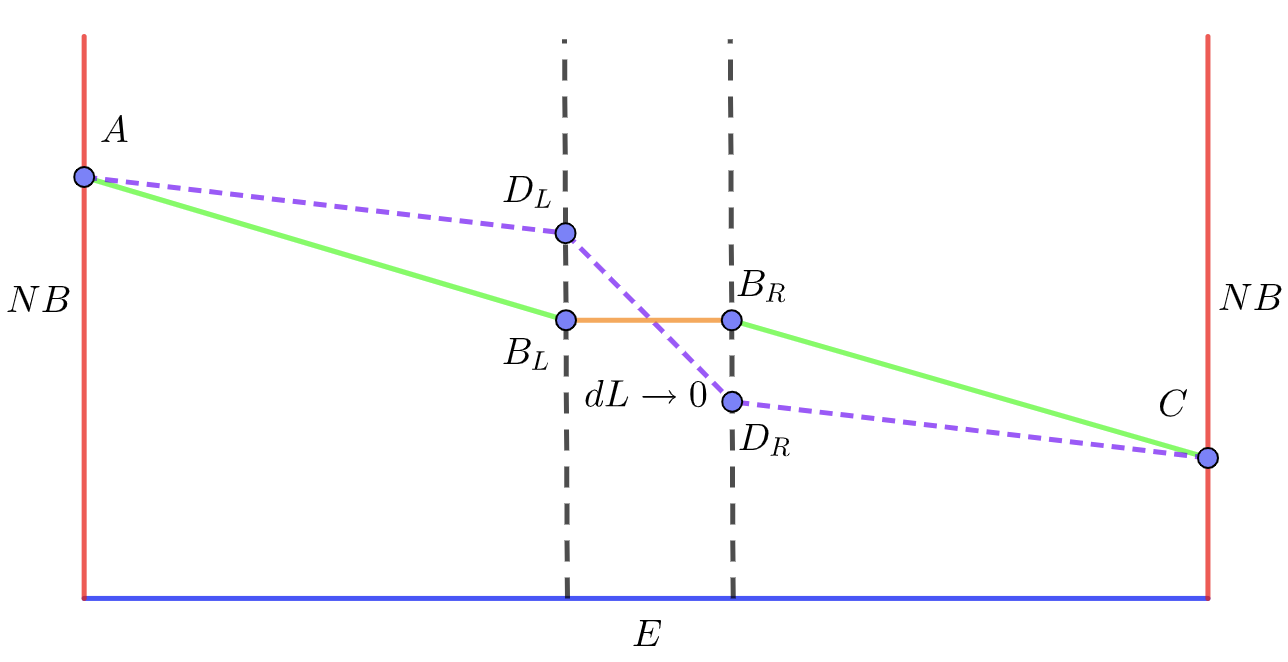}
    \caption{When we increase the edge length by $dL \to 0$ (represented by the orange line), the RT surface transitions from the green curves $AB_L \cup B_RC$ to the purple dashed curves $AD_L \cup D_L D_R \cup D_RC$. }
    \label{fig:deformedRT}
\end{figure}

We will demonstrate that the HEE increases when the internal edge is slightly extended by a length $dL$. In Fig. \ref{fig:deformedRT}, the change is illustrated, focusing on Poincaré AdS for simplicity. When we increase the edge length by $dL \to 0$ (represented by the orange line), the RT surface transitions from the green curves $AB_L \cup B_RC$ to the purple dashed curves $AD_L \cup D_L D_R \cup D_RC$. Since $dL$ is very small, the local region around $ D_L D_R$ can be treated as flat space if we ignore second-order corrections $O(dL^2)$. Thus, we can write
\begin{align}\label{sect4: prove dL 1}
\text{Area} (D_LD_R)>\text{Area} ( D_LB_L\cup D_RB_R),
\end{align} 
in the linear order of $dL$. Similarly, we have near the local regions of $D_LB_L$ and $D_RB_R$ that
\begin{align}\label{sect4: prove dL 2}
\text{Area} (AD_L\cup D_L B_L)>\text{Area} ( AB_L),\ \ \text{Area} (CD_R\cup D_RB_R )>\text{Area} ( CB_R).
\end{align} 
Combining these inequalities, we find
\begin{align}\label{sect4: prove dL 3}
\text{Area} (AD_L\cup D_L D_R \cup D_RC)&>\text{Area} (AD_L \cup D_RC)+\text{Area} ( D_LB_L\cup D_RB_R)\nonumber\\
&> \text{Area} (AB_L\cup B_RC),
\end{align} 
which confirms that the HEE indeed increases with an increase in internal edge length.

From now on, we will focus on the correct proposal for HEE involving connected RT surfaces on the Net-brane. We aim to analyze how these RT surfaces are interconnected on a single Net-brane. We obtain the following connecting condition
 \begin{align}\label{sect4: connecting condition RT}
\sum_{m=1}^p \overset{(m)}{n}{}^{\alpha}\frac{\partial x^{\mu}}{\partial \xi^{\alpha}} \frac{\partial x^{\nu}}{\partial y^i} g_{\mu\nu}|_{\Gamma\cap \text{NB}}=0,
\end{align}
where $\overset{(m)}{n}{}^{\alpha}$ are the unit vectors normal to the intersection $\gamma=\Gamma\cap NB$ directed along the RT surface $\Gamma$ of branch $B_m$ approaching the Net-brane $NB$, $x^{\mu}$, $\xi^{\alpha}$ and $y^i$ are the coordinates of the bulk, the RT surfaces $\Gamma$ and Net-brane $NB$, respectively. It is important to note that the AdS/NCFT simplifies to AdS/BCFT under the $Z_p$ symmetry. Under this condition, the connecting condition (\ref{sect4: connecting condition RT}) reduces to the orthogonality condition for the RT surface and EOW brane in AdS/BCFT. It can be regarded as a test of our result (\ref{sect4: connecting condition RT}). 

Similar to the junction condition for bulk branches, the connecting condition for RT surfaces can be derived from the variational principle. See \cite{Chen:2020uac} for a similar case in AdS/ICFT.  We label the embedding function of RT surfaces as $x^{\mu}(\xi^{\alpha})$ and get the induced metric on the RT surface $\Gamma$ as follows
\begin{align}\label{sect4: RT metric}
h_{\Gamma \ \alpha\beta}=\frac{\partial x^{\mu}}{\partial \xi^{\alpha}} \frac{\partial x^{\nu}}{\partial \xi^{\beta}} g_{\mu\nu}.
\end{align}
The area of the RT surface is given by
\begin{align}\label{sect4: RT area}
A=\int_{\Gamma}d^{d-1}\xi \sqrt{h_{\Gamma}}. 
\end{align}
Taking variations of this area functional with respect to $x^{\mu}$ and integrating by parts, we obtain
\begin{align}\label{sect4: variation RT area}
\delta A=-\int_{\Gamma}d^{d-1}\xi \sqrt{h_{\Gamma}} K_{\Gamma\ \mu} \delta x^{\mu}+\int_{\gamma}d^{d-2}y \sqrt{h_{\gamma}}\sum_m \overset{(m)}{n}{}^{\alpha}\frac{\partial x^{\mu}}{\partial \xi^{\alpha}} g_{\mu\nu}\delta x^{\nu}=0,
\end{align}
where $K_{\Gamma\ \mu} $ represents the trace of the extrinsic curvatures from the bulk to the RT surface, and $\gamma = \Gamma \cap \text{NB}$ denotes the intersection between the RT surface $\Gamma$ and the Net-brane $NB$. The extrinsic curvature is defined by
\begin{align}\label{sect4: Kuij}
K_{\Gamma}{}^{\mu}_{\alpha\beta}= \partial_{\alpha}\partial_{\beta} x^{\mu}-\gamma^{\gamma}_{\alpha\beta} \partial_{\gamma} x^{\mu}+\Gamma^{\mu}_{\sigma\rho}\partial_{\alpha} x^{\sigma}\partial_{\beta} x^{\rho},
\end{align}
where $\gamma^{\gamma}_{\alpha\beta}$ and $\Gamma^{\mu}_{\sigma\rho}$ are the Levi-Civita connections with respect to $h_{\alpha\beta}$ and $g_{\mu\nu}$, respectively. The bulk term of (\ref{sect4: variation RT area}) indicates that the RT surface is a minimal surface with  
\begin{align}\label{sect4.1: minimal surface}
K_{\Gamma\ \mu} =0.
\end{align}
Furthermore, since $ \delta x^{\mu}|_{\gamma} =( \partial x^{\mu}/\partial y^i ) \delta y^i$ is defined along the Net-brane, the boundary term of (\ref{sect4: variation RT area}) gives us the connecting condition (\ref{sect4: connecting condition RT}) for the RT surfaces.

\subsection{Network entropy}

This subsection discusses holographic entanglement entropy for networks. Let us start with the simplest network, which has $p$ edges linked by a single node. The vacuum of this network is dual to Poincaré AdS (\ref{sect2: AdS metric},\ref{sect2: AdS Nbrane}). For simplicity, we focus on AdS$_3$ and choose the subsystem as $A: \overset{(m)}{x} \le L_m$. The embedding function of RT surface reads \cite{Ryu:2006bv}
\begin{align}\label{sect4: embedding function RT}
z^2+( \overset{(m)}{x}-\overset{(m)}{a})^2=\overset{(m)}{b}{}^2,
\end{align}
where $\overset{(m)}{a}, \overset{(m)}{b}$ are free parameters. Substituting typical points $(0, L_m)$ and $(z_b, x_b)$ into the above embedding function, we determine the parameters $\overset{(m)}{a}, \overset{(m)}{b}$ and drive the area of RT surface
\begin{align}\label{sect4: minimal surface}
A=\sum_{m=1}^p\log \left(\frac{(L_m-x_b)^2+z_b^2}{z_b \epsilon }\right),
\end{align}
$\epsilon$ labels the UV cut-off of $z$, and $(z_b, x_b)$ denotes the common intersection point of RT surfaces on the Net-brane. According to (\ref{sect2: AdS Nbrane}), we have $x_b=-\sinh(\rho) z_b$. By minimizing the area (\ref{sect4: minimal surface}), we can determine $z_b$. For the symmetric case $L_m=L$, we have $z_b=L \text{sech}(\rho )$, yielding 
\begin{align}\label{sect4: minimal surface1 }
A|_{L_m=L}=p\log \left(\frac{2L}{ \epsilon }\right)+p \rho.
\end{align}
From (\ref{sect4: minimal surface}), we read off the universal logarithmic divergent term of entanglement entropy
\begin{align}\label{sect4: divergent term}
S_{\text{NCFT}}=\frac{A}{4G_N}= \frac{c p}{6} \log \left(\frac{1}{\epsilon }\right)+{\text{finite terms}},
\end{align}
where $c=3l/2G_N$ is the central charge, $p$ is the number of edges. Note that we have set the AdS radius $l=1$ in AdS$_3$.

\begin{figure}[htbp]
  \centering
\includegraphics[width=.6\textwidth]{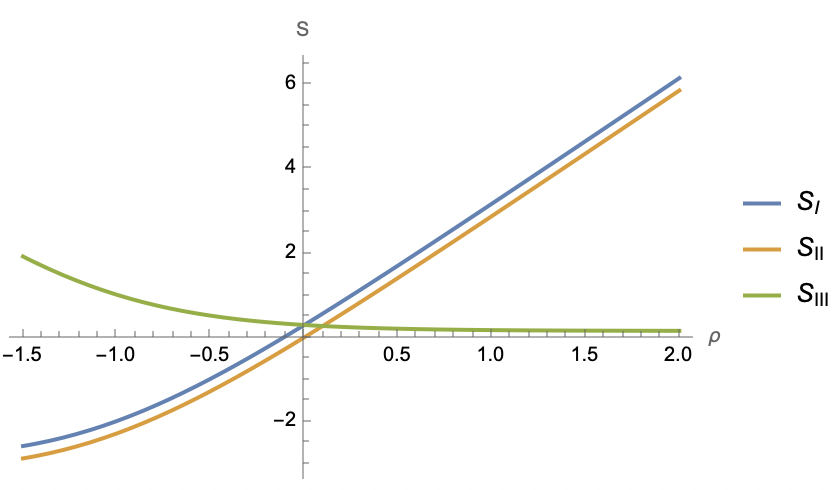}
 \caption{Various network entropies vary with the brane tension $T=p(d-1)\tanh\rho$. We choose $L_1=1, L_2=2, L_3=3$ and set $c/6=1$. The blue, orange, and green curves denote $S_{\text{I}}$, $S_{\text{II}}$ and $S_{\text{III}}$, respectively. It shows $S_{\text{I}}$ and $S_{\text{II}}$ increase with $\rho$, while $S_{\text{III}}$ decreases with $\rho$. Besides, $S_{\text{I}}=S_{\text{III}}$ at $\rho=0$, $S_{\text{I}}\ge S_{\text{II}}$ and $S_{\text{III}}\ge 0$. } \label{networkentropy}
\end{figure}

Let us go on to analyze the finite terms, which include the network information. There are several natural ways to extract the network contributions to entanglement entropy. We define the type I network entropy as the difference between the entanglement entropy of NCFT and CFT
\begin{align}\label{sect4: type I S}
S_{\text{I}}=S_{\text{NCFT}}-S_{\text{CFT}},
\end{align}
where $S_{\text{CFT}}$ is half the entanglement entropy of a strip with the width $2L_m$
\begin{align}\label{sect4: S CFT}
S_{\text{CFT}}=\frac{c}{6}\sum_{m=1}^p \log \left(\frac{2L_m}{ \epsilon }\right).
\end{align}
For the symmetric case $L_m=L$, we get 
\begin{align}\label{sect4: S CFT}
S_{\text{I}}|_{L_m=L}=\frac{c p}{6}\rho,
\end{align}
which increases with the brane tension (\ref{sect2: AdS brane tension}). According to \cite{Fujita:2011fp}, we have $\rho_{\text{UV}}\ge \rho_{\text{IR}}$ under boundary RG flow. The case for the network node is similar. Thus, similar to the boundary entropy, the type I network entropy decreases under the renormalization group (RG) flow  
\begin{align}\label{sect4: type I RG flow}
S_{\text{I}}|_{\text{IR}} \le S_{\text{I}}|_{\text{UV}}. 
\end{align}
It should be stressed that (\ref{sect4: type I RG flow}) also holds for the asymmetric case $L_i\ne L_j$. See Fig. \ref{networkentropy} for an example. 

Inspired by the boundary entropy in AdS/BCFT, we propose another natural definition of network entropy
\begin{align}\label{sect4: type II S abc}
S_{\text{II}}=S_{\text{NCFT}}(\rho)-S_{\text{NCFT}}(0),
\end{align}
which we refer to as type II network entropy. Note that $S_{\text{CFT}}$ represents the disconnected HEE proposal of sect. 4.1 with $\rho=0$, which is smaller than the connected HEE proposal $S_{\text{NCFT}}(\rho = 0)$. Consequently, we have the following inequality
\begin{align}\label{sect4: type II S}
S_{\text{II}}\le S_{\text{I}},
\end{align}
where this inequality is saturated in the symmetric case $L_m = L$. Similar to $S_{\text{I}}$, the type II network entropy $S_{\text{II}}$ increases with the brane tension, adhering to the RG flow relation
\begin{align}\label{sect4: type II RG flow}
S_{\text{II}}|_{\text{IR}} \le S_{\text{II}}|_{\text{UV}}. 
\end{align}
See Fig. \ref{networkentropy} for an example.

The entropy difference between NCFT and BCFT defines the type III network entropy
\begin{align}\label{sect4: type III S}
S_{\text{III}}=S_{\text{NCFT}}-S_{\text{BCFT}},
\end{align}
where $S_{\text{BCFT}}$ is given by the area of extremal surfaces normal to the Net-brane
\begin{align}\label{sect4: S BCFT}
S_{\text{BCFT}}=\frac{cp}{6} \rho+\frac{c}{6}\sum_{m=1}^p \log \left(\frac{2L_m}{ \epsilon }\right).
\end{align}
It is the first proposal of HEE discussed in sect.4.1. Since $S_{\text{BCFT}}$ is defined by the area of minimal surfaces ending on the Net-brane, not necessarily intersecting at a single point as $S_{\text{EE}}$, we have $ S_{\text{BCFT}} \le S_{\text{NCFT}}$. As a result, the type III network entropy is semi-positive definite
\begin{align}\label{sect4: S BCFT 1}
S_{\text{III}}\ge 0. 
\end{align}
Note that $S_{\text{III}}=0$ for the symmetric case $L_m=L$. Let us discuss the physical meanings of type III network entropy further. As discussed in sect. 2.3, the spectrum of NCFTs consists of isolated modes (AdS/BCFT with NBC) and transparent modes (AdS/BCFT with DBC/CBC). The entanglement entropy of isolated modes at each edge is represented by $S_{\text{BCFT}}$.
In contrast, $S_{\text{NCFT}}$ accounts for the entanglement among all modes in NCFTs, including transparent modes, isolated modes, and the correlations between them. Thus, $S_{\text{III}}$ characterizes the entanglement between NCFT modes, excluding those from isolated modes. Given this interpretation, it follows naturally that $S_{\text{III}}$ is non-negative. 
According to \cite{Pang:2025flq}, there is more than one choice of junction condition (JC) for NCFTs, similar to boundary conditions (BC) for BCFTs. Each type of JC in NCFTs corresponds to a specific BC in BCFTs. For instance, the JC (\ref{sect3: JC1 scalar}) corresponds to the Neumann boundary condition of scalar fields. It can be demonstrated by setting $p=1$ in JC (\ref{sect3: JC1 scalar}). Therefore, for the NCFTs with JC X, $S_{\text{BCFT}}$ of (\ref{sect4: type III S}) reflects the entropy for BCFTs with BC X. We plan to compute $S_{\text{III}}$ for free BCFTs in our upcoming research. Should it also indicate a positive $S_{\text{III}}$, it would further substantiate our interpretations of $S_{\text{III}}$, excluding the entanglement entropy of isolated modes at each single edge.

\begin{figure}[htbp]
  \centering
\includegraphics[width=.6\textwidth]{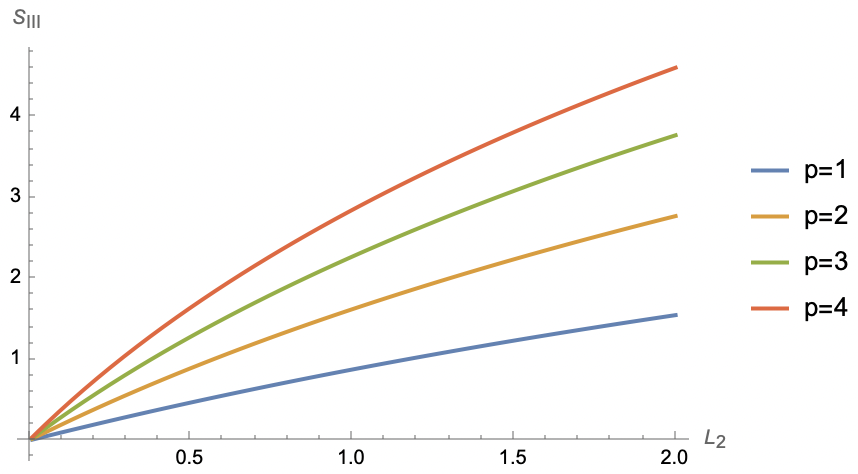}
 \caption{The type III network entropy increases with the length $L_2$ and number $p$ of internal edges. Here we have set the length of the external edge $L_1=1$ and $c/6=1$. The blue, orange, green and red curves correspond to $p=1,2,3,4$, respectively. } \label{networkentropy III}
\end{figure}

Unlike $S_{\text{I}}$ and $S_{\text{II}}$, the type III network entropy $ S_{\text{III}}$ decreases with increasing $\rho$ (see Fig. \ref{networkentropy}). Thus, it does not reflect the degrees of freedom at the nodes. Instead, it effectively describes the network's structure and size, including the lengths and the number of internal edges. To illustrate this, we consider Poincaré AdS$_3$ with tensionless Net-branes as an example. For a network consisting of two nodes, four external edges, and $p$ internal edges, the expression for the type III network entropy $S_{\text{III}}$ is given by:
\begin{align}\label{sect4: S3 increase with network size}
S_{\text{III}}=\frac{c}{6} \Big[ 2 p \tanh ^{-1}\left(\sqrt{\frac{L_2^2}{4 z_b^2+L_2^2}}\right)+4 \log \left(\frac{z_b^2+L_1^2}{2 L_1 z_b}\right)  \Big],
\end{align}
where $L_1$ and $L_2$ are the lengths of connected subsystem $A$ on the external and internal edges, respectively. The intersection $z_b$ of the RT surfaces on the Net-brane can be determined numerically by minimizing the equation above. We have numerically verified that $S_{\text{III}}$ increases with the length $L_2$ and the number $p$ of internal edges (see Fig. \ref{networkentropy III}). As discussed around eqs. (\ref{sect4: prove second entropy},\ref{sect4: prove dL 3}), $S_{\text{NCFT}}$ increases with the lengths and 
number of internal edges, while $S_{\text{BCFT}}$ is independent of the internal edges. Consequently, the monotonic increase of $S_{\text{III}}$ concerning edge length and number always holds.

To end this section, we draw various network entropies in Fig. \ref{networkentropy}. It shows $S_{\text{I}}$ and $S_{\text{II}}$ increase with brane tension, while $S_{\text{III}}$ decreases with brane tension. Besides, $S_{\text{I}}=S_{\text{III}}$ at $\rho=0$, $S_{\text{I}}\ge S_{\text{II}}$ and $S_{\text{III}}\ge 0$.
 
\section{Holographic network problems}
\label{sect for HNP}

This section provides a brief discussion of famous network and graph problems from a holographic perspective. We use the shortest path problem as an example and reserve the study of other famous network and graph problems for future work. The shortest path problem seeks to determine the shortest path from a starting point to an ending point within a network. For example, calculate the shortest driving route in the urban road network from one place to another. Dijkstra's algorithm is a classic algorithm for solving this problem. 
In the following, we first construct the gravity dual of the shortest path problem, then show that the shortest path on the network is dual to the shortest geodesic in the bulk, and finally discuss its connection to holographic two-point correlators for massive operators.

Consider a one-dimensional Euclidean network with $n$ nodes $N$ and $p$ edges $E_m$ of length $L_m$. Each edge $E_m$ is dual to an Euclidean AdS$_2$
\begin{align}\label{sect5: AdS2}
d\overset{(m)}{s}{}^2=\frac{dz^2+d\overset{(m)}{x}{}^2}{z^2}, \ 0 \le \overset{(m)}{x}\le L_m,
\end{align}
where the nodes are located at $z=0, \overset{(m)}{x}=0, L_m$. We glue the branch $B_m$ with tensionless Net-branes to get the gravity dual of the network.
While the Poincaré AdS (\ref{sect5: AdS2}) may not represent the vacuum state, it does not impact our goal. Our objective is to identify a gravity dual for the short-path problem, which doesn't have to be a vacuum solution. As we will demonstrate below, the glued Euclidean AdS$_2$ with the tensionless Net-branes suits our needs.

The network generally allows multiple paths from the starting point $A$ to the endpoint $B$. The paths must not contain loops to ensure they are the shortest routes. Consider an arbitrary path $\mathcal{P}_l$ that traverses the edge $\mathcal{E}_l=\{ E_i| E_i\cap \mathcal{P}_{l}\ne \varnothing \} $.  This edge set $\mathcal{E}_l$ selects a branch set $\mathcal{B}_l=\{ B_i| B_i\cap \mathcal{E}_l \ne \varnothing \}$ in the bulk. Since the path $\mathcal{P}_l$ is loop-free, the branch set  $\mathcal{B}_l$ is also loop-free. In other words, each branch connects to at most one branch on one Net-brane in the set $ \mathcal{B}_l$. Since the Net-branes are tensionless, the branch set \( \mathcal{B}_l \) can be viewed as a smooth AdS\(_2\) spacetime without Net-branes. The geodesic distance connecting points A and B in AdS$_2$ is given by
\begin{align}\label{sect5: AdS2 geodesic}
L_{AB}=2 \log(\frac{l_{AB}}{\epsilon}),
\end{align}
which increases monotonically with the distance $l_{AB}$ between points $A$ and $B$ in $\mathcal{E}_l$. Here $\epsilon$ is the UV cut-off of $z$. Thus, minimizing the distance $l_{AB}$ on the network is equivalent to minimizing the distance $L_{AB}$ in the bulk.

In AdS/CFT, the two-point function of local operators dual to massive scalars is determined by the proper distance in the bulk
\begin{align}\label{sect5: AdS2 two point function}
\langle  O(A) O(B) \rangle \sim e^{-M L_{AB}},
\end{align}
where $M\gg 1$ denotes the scalar mass. Let us give a quick review of its derivation in AdS/CFT. For our purpose, we focus on the massive scalar in the probe limit of Euclidean AdS$_2$, where the two-point function of the dual operator $O$ is given by \cite{Witten:1998qj}
\begin{align}\label{sect5: OO 1}
\langle  O(A) O(B) \rangle \sim \frac{1}{l_{AB}^{2\Delta}},
\end{align}
where $l_{AB}=|x_A-x_B|$ is the distance between points A and B in a flat space, and the conformal dimension is calculated as  \cite{Witten:1998qj}
\begin{align}\label{sect5: OO 2}
\Delta=\frac{1}{2}(1+\sqrt{1 + 4M^2}) \sim M. 
\end{align}
By using (\ref{sect5: AdS2 geodesic}, \ref{sect5: OO 2}) in the large $M$ limit, we can reformulate (\ref{sect5: OO 1}) as (\ref{sect5: AdS2 two point function}).

Let us now generalize the discussions to the multi-Euclidean AdS$_2$ glued by tensionless Net-branes. We follow the approach of \cite{Witten:1998qj} and take into account the effects of junction conditions on the Net-brane. We find that the two-point functions of scalar operators in networks can be expressed as follows:
\begin{align}\label{sect5: OO 3}
\langle  O(A) O(B) \rangle =\sum_{\text{all path }\mathcal{P}_{i}} c_i \frac{1}{(l_{i\ AB})^{2\Delta}},
\end{align}
where $l_{i\ AB}$ denotes the length of boundary path $\mathcal{P}_{i}$
connecting points $A$ and $B$ in the networks. The constants $c_i$ depend on the reflection and transmission coefficients $c_r, c_t$ of the nodes along the path, and the summation is over all possible paths, including those with loops. Refer to Appendix \ref{app for two point function} for the derivation of (\ref{sect5:  OO 3}). In the limit of large mass, where $\Delta \sim M \to \infty$, the shortest path dominates the correlator (\ref{sect5: OO 3}). By utilizing (\ref{sect5: AdS2 geodesic}), we finally obtain the following result in the large mass limit
\begin{align}\label{sect5: OO 4}
\langle  O(A) O(B) \rangle \sim e^{-M L_{\text{min}\ AB}},
\end{align}
where $L_{\text{min}\ AB}$ denotes the length of shortest bulk path connecting two points $A,B$ in networks. Thus, the shortest path problem of the network is equivalent to calculating the two-point correlators (\ref{sect5: OO 4}) in the large mass or dimension limit \footnote{We assume $L_{\text{min}\ AB}$ is still a monotonically increasing function of $l_{AB}$ even in the spacetime with back-reactions due to massive scalars. Otherwise, the two-point correlator (\ref{sect5: AdS2 two point function}) could increase with $l_{AB}$, which is unphysical.}.

We consider a variation of the shortest path problem where nodes in the network have weights, such as toll stations. Each time you pass through a toll station $N_i$, you incur a cost of $C_i$. The goal is to find the most economical route from point A to point B, minimizing the total price, including fuel and toll fees. The total cost can be expressed as
$(\lambda l_{AB}+\sum_{N_i \in \mathcal{P}_l } C_i)$, where $\lambda$ is the fuel cost per distance, $ l_{AB}$ is the length of the journey, and $\mathcal{P}_l$ is the path from A to B. To simplify this problem, we can convert it into a standard shortest path problem by redefining the length between two nodes $N_i$ and $N_j$: $\bar{L}_{ij} = L_{ij} + \frac{C_i + C_j}{2\lambda}$. This way, minimizing the new length $\bar{l}_{AB}$ will also minimize the total costs effectively.

In summary, we examined the holographic shortest path problem for networks and its variant. It is equivalent to finding the shortest distance in bulk or the holographic two-point correlators for massive operators. While the holographic approach does not simplify the original shortest path problem, it could offer new insights into specific network issues. We will explore this intriguing problem in future research.

\section{Conclusions and Discussions}

This paper explores the network conformal field theory (NCFT) and its gravitational dual. We suggest that a network node connecting $p$ edges is dual to a Net-brane linking $p$ branches in the bulk spacetime. We establish junction conditions for the Net-brane, demonstrating that they lead to energy conservation on the network node, which strongly supports our proposal of AdS/NCFT. We examine the typical geometries of the holographic network, notably finding that Poincaré AdS and AdS black holes serve as gravitational duals for various networks. It allows us to extend many results from AdS/CFT to AdS/NCFT. We remark that Poincaré AdS does not represent the vacuum state of a network with internal edges due to a non-trivial Casimir effect. We find that the spectrum of gravitational KK modes on the Net-brane is a mixture of that of AdS/BCFT with NBC and DBC/CBC, corresponding to the isolated and transparent modes, respectively. We also investigate two-point correlation functions in NCFTs, providing examples that include free theories and AdS/NCFT with a tensionless Net-brane.

We propose that the RT surface connects to the same intersection on the Net-brane for a connected subsystem within the network, and we validate this with the strong additivity and monotonicity of entanglement entropy. Additionally, we derive connecting conditions for the RT surfaces on the Net-brane, showing they align with the orthogonality condition for the RT surface and EOW brane in AdS/BCFT under the $Z_p$ symmetry. We present various definitions of network entropy, with one based on the difference in entanglement entropy between NCFT and BCFT, which is always non-negative and effectively captures the network's complexity. Finally, we briefly discuss the holographic perspective of the shortest path problem. We find it is closely related to the shortest geodesic in bulk and the holographic two-point correlators of massive operators.

Many interesting problems are worthy of exploration.

\begin{itemize}

 \item This paper primarily focuses on simple networks. Exploring complex networks is highly intriguing, as quantum entanglement and correlation are anticipated to reveal more intricate structures and novel properties. Deep learning serves as a powerful tool for analyzing physics in complex networks. See \cite{Ahn:2024gjf, Chen:2024ckb, Cai:2024eqa,Halverson:2024axc} for the applications of deep/machine learning in holography and CFTs. 
 
 \item Characterizing the complexity of networks is a significant question. Deep learning suggests that artificial intelligence can only emerge when networks reach a certain level of complexity (the network is ``deep" enough) \footnote{By ``complexity", we mean the network's structural complexity. Please do not confuse it with the holographic complexity related to the evolution of eternal black holes \cite{Brown:2015bva,Belin:2021bga}.}. The key question is: What is the threshold for this complexity? Additionally, how does this critical complexity manifest from a holographic perspective? These are essential questions of broad interest. 
 
 \item This paper only discusses the HEE of a network. Studying the entanglement entropy of free theories on the network is interesting. We argue that the network entropy $S_{\text{III}}$ is non-negative and verify it with AdS/NCFT. It is interesting to further test $S_{\text{III}}\ge 0$ with free theories. 
 
 \item As mentioned earlier, Poincaré AdS is not dual to the vacuum state of a general network. It is crucial to find the gravitational dual of the vacuum state and examine the holographic bound of the Casimir effect \cite{Miao:2024gcq, Miao:2025utb} for various networks. 
  
 \item For simplicity, we assume the same CFTs on different edges, corresponding to the same gravity theory in different bulk branches. Studying the AdS/NCFT with different gravity parameters or theories in various branches is interesting.  
 
 \item This paper focuses on the junction condition that requires the field to be continuous, while the sum of the normal derivatives of the field must vanish at the node or Net-brane. It corresponds to Neumann boundary conditions (NBC) for a BCFT with one edge (p = 1). Similar to boundary conditions, there are alternative junction conditions that necessitate the continuity of the normal derivative of the field, while the sum of the fields at the node or Net-brane equals zero \cite{Pang:2025flq}. This new junction condition corresponds to Dirichlet boundary conditions (DBC) in a BCFT. It is interesting to investigate the correlation functions and entanglement entropy associated with this new type of junction condition. Similar to the AdS/BCFT with DBC, AdS/NCFT with this new junction condition on the Net-brane is expected to be dual to open networks, where the total current or energy flowing into the node is non-zero. We will address a detailed study of this issue in future work.

 \item Last but not least, our framework can be used to study holographic circuits and predict various electrical issues. We can construct holographic circuits by gluing geometries, such as AdS black holes, with various bulk fields corresponding to resistors, capacitors, inductors, diodes, etc. Does holography establish a limit on chip computing speed or the transport coefficient between different branches? These questions are worth exploring.
 
 \end{itemize}

\acknowledgments

Miao thanks S. He for the valuable discussions during the 30th anniversary of AEI. Miao acknowledges the supports from National Natural Science Foundation of China (NSFC) grant (No.12275366).

 \paragraph{Note added.}  After we have finished the work, a paper appears in arXiv \cite{Liu:2025khw} with some overlaps with our paper. They focus on the holographic energy transport for 2d CFTs on intersecting lines. In comparison, we explore the gravity dual of CFTs in general networks in general dimensions.

\appendix

\section{Conventions and Notations}\label{app for notations}

In this appendix, we provide a summary of the notations used in this paper in Table \ref{app: conenotation}. 

\begin{table}[ht]
\caption{Notations of AdS/NCFT}
\begin{center}
    \begin{tabular}{| c | c | c | c |  c | c | c | c| c| c|c|c| }
    \hline
    & network edge $E_m$ & network node $N$  & bulk branch $B_m$ & Net-brane $NB$  \\ \hline
 coordinate & $y^i\text{ or }x^{i}=(\overset{(m)}{x}, y^{a})$ & $y^{a}$ & $x^{\mu}=(z,\overset{(m)}{x}, y^{a})$ & $y^i=(z, y^{a})$ \\ \hline
 metric& $h_{ij},\ h_{E\ ij}$ & $\sigma_{ab}$ & $g_{\mu\nu}$ &$h_{ij},\ h_{NB\ ij}$\\ \hline
 normal direction& $\hat{n}$ & $n$ & $\varnothing$ & $\hat{n}$ \\ \hline
 \end{tabular}
\end{center}
\label{app: conenotation}
\end{table}

Let us explain some key points for the notations. The edges \(E_m\) are situated at the AdS boundary (\(z=0\)), while the node \(N\) is at \(\overset{(m)}{x}=z=0\). The Net-brane \(NB\) is positioned at \(\overset{(m)}{x}=-\sinh(\rho) z\), with \(\rho\) representing a parameter related to the brane tension. In this paper, we focus on the scenario where all bulk branches \(B_m\) share the same AdS radius, \(l_m=1\). More general cases will be discussed in a future paper. The notation \(h_{ij}\) refers to the induced metric on either the network edges \(E_m\) or the Net-brane \(NB\). It's used consistently in expressions like \(\int_{E_m} h_{ij}\), \(h_{ij}|_{E_m}\), or \(\int_{NB} h_{ij}\), \(h_{ij}|_{NB}\), which eliminates ambiguity. Similarly, the coordinates \(y^i\) and the normal vectors \(\hat{n}\) and \(n\) are defined clearly. Here, \(\hat{n}\) points from the bulk toward the edges or the Net-brane, while \(n\) points from the edges or the Net-brane toward the node. Additionally, the vector \(\overset{(m)}{n}{}^{\alpha}\) in Eq.~(\ref{sect4: connecting condition RT}) is the unit normal vector to the intersection \(\gamma=\Gamma_m\cap NB\), defined along the RT surface \(\Gamma_m\) as it approaches the Net-brane.

In section \ref{section 2.1}, we have utilized both the coordinates outlined in Table~\ref{app: conenotation} and additional coordinate systems, specifically Gaussian normal coordinates and the Fefferman–Graham gauge, to derive the energy conservation law at the node based on the junction condition on the Net-brane. Let us clarify these coordinate systems. We adopt Gaussian normal coordinates \(y^{i}=(l,y^{a})\) for both the edges and the Net-brane in the vicinity of the node. Here, \(l\) parametrizes the normal direction pointing from the node into the interior of the Net-brane or the edges, with \(l=0\) indicating the location of the node. For further discussion on Gaussian normal coordinates, please refer to Appendix \ref{app for Killing vectors}. Additionally, to transform the extrinsic curvature on the AdS boundary into the CFT stress tensor \(T^{\text{CFT}}_{ij}\), we introduce another coordinate system in the Fefferman–Graham gauge, defined by the coordinates \(x^{i}=(x,x^{a})\) on regularized edges at \(z=\epsilon\). The transformations between these two coordinate systems are given by Eq.~(\ref{sect2: GN coordinates}).

In section \ref{sect for correlation function}, we designate the coordinates on each edge \(E_m\) as \(\overset{(m)}{\mathbf{x}}\), where \(\overset{(m)}{x}{}^{i} = (\overset{(m)}{x}, \overset{(m)}{y}{}^{a})\) when discussing the two-point functions. It is important to note that \(\overset{(m)}{\mathbf{x}}\) refers to the coordinate vector \(\overset{(m)}{x}{}^{i}\), which should be distinguished from the coordinate \(\overset{(m)}{x}\) along the edge \(E_m\), representing the distance from the node. Furthermore, in the general case that involves both points on the same edge and on different edges, we will omit the upper index \((m)\) for simplicity. For instance, consider \(\overset{(m)}{T}_{ij}(\mathbf{x})\); the argument \(\mathbf{x}\) should be understood as \(\overset{(m)}{\mathbf{x}}\), specifically the coordinates on the edge \(E_m\).

In section \ref{sect for HEE}, we derive the connecting condition for the RT surface at its intersection with the Net-brane. We summarize the related notations in Table \ref{app: RT notation}.

\begin{table}[!h]
\caption{Notations about connecting condition}
\begin{center}
    \begin{tabular}{| c | c | c |  c | c | c | c| c| c|c|c| }
    \hline
     & bulk branch $B_m$ &Net-brane $NB$& RT surface $\Gamma$& intersection $\gamma = \Gamma\cap NB$  \\ \hline
 coordinate &  $x^{\mu}$ &$y^{i}$& $\xi^{\alpha}$ & $y^{a}$ \\ \hline
 metric& $g_{\mu\nu}$ &$h_{ij}$& $h_{\Gamma~\alpha\beta}$ &$ h_{\gamma~ab}$\\ \hline
 \end{tabular}
\end{center}
\label{app: RT notation}
\end{table}

\section{Killing vectors at linear order}\label{app for Killing vectors}

There are no exact Killing vectors present in an arbitrary spacetime. Even locally, we can at most get a Killing vector at the linear order of coordinates. Take the network node $N$ as an example. Near a local point on $N$, we can expand the metric in Riemann normal coordinates 
\begin{align}\label{app A: N metric}
ds_N^2=\sigma_{ab}dy^a dy^b=\Big(\eta_{ab}-\frac{1}{3} R_{Nacbd}\  y^c y^d+O(y^3)\Big)dy^a dy^b,
\end{align}
where $R_{Nacbd}$ are curvatures at the origin $y^a=0$. Solving the Killing equations $\bar{D}_{(a}\bar{\xi}_{b)}=0$, we get $(d-1)$ local Killing vectors at the first order
\begin{align}\label{app A: N Killing}
\bar{\xi}^a=\delta^a_{\ b}+O(y^2),
\end{align}
where $b=1,2,...,d-1$. In general, no Killing vectors exist at the next order $O(y^2)$, unless we impose constraints on the curvatures $R_{Nacbd}$. 

Let us go on to discuss a local region $dV=dl dS$ including a small piece $dS$ of the node $N$ in the Net-brane $NB$ or the network edges $E_m$. We adopt the Gauss normal coordinates $y^i=(l, y^a)$
\begin{align}\label{app A: NB metric}
ds^2&=h_{ij}dy^i dy^j=dl^2+\Big( \sigma_{ab} -2l k_{ab}(y)+O(l^2)\Big) dy^a dy^b\nonumber\\
&=dl^2+\Big( \eta_{ab} -2l k_{ab}(0)+O(l^2, y^2, l y)\Big) dy^a dy^b,
\end{align} 
where $k_{ab}$ are extrinsic curvatures on the node. We focus on flat network edges $E_m$ with flat boundaries so that $k_{ab}=0$ for edges $E_m$. On the other hand, the Net-brane is an asymptotically AdS spacetime and $k_{ab}\ne 0$ on its boundary generally. 
Note that the traceless components $\bar{k}_{ab}=0$ vanish on the asymptotically AdS boundary. 

Solving the Killing equations $D_{(i}\xi_{j)}=0$ perturabtively, we obtain $(d-1)$ local Killing vectors $\xi^i=(\xi^l, \xi^a)$ with 
\begin{align}\label{app A: NB Killing}
\xi^l=O(y^i)^2, \  \xi^a=\delta^a_{\ b}- 2 l k^{a}_{\ b}(0) +O(y^i)^2,
\end{align}
where $b=1,2,...,d-1$. The above Killing vectors obey $D_{(i}\xi_{j)}=O(y^i)$ and $n_i \xi^i=O(y^i)^2$, where $n_i=(-1,0,..,0)$ is the normal vector on the node $l=0$. 

\section{Holographic one-point function of scalar operator}
\label{app for holo one-point function}

This appendix studies the one-point function of the operator dual to the bulk scalar. For simplicity, we focus on a massless bulk scalar in the probe limit of AdS$_5$, described by the metric:
\begin{align}\label{app2: AdS metric}
ds^2=\frac{dz^2+d\overset{(m)}{x}{}^2+\delta_{ab} dy^ady^b}{z^2},
\end{align}
with the node at $z=\overset{(m)}{x}=0$ and the Net-brane at 
\begin{align}\label{app2: AdS Nbrane}
\text{Net-brane}:\ \overset{(m)}{x}=-\sinh(\rho)z.
\end{align}
As discussed in sect. 2, we impose the following junction condition on the Net-brane
\begin{align} \label{app2: JC1 scalar bulk}
 \text{JC for bulk scalar: }\ \ \ \overset{(m)}{ \Phi}|_{NB}=\overset{(n)}{\Phi}|_{NB}, \ \ \  \sum_{m=1}^p\partial_{\hat{n}}\overset{(m)}{\Phi}|_{NB}=0.
\end{align}

Following \cite{Chu:2020gwq}, we use the ansatz for the bulk scalar on branch $B_m$:
\begin{align} \label{app2: bulk scalar}
\overset{(m)}{\Phi}= \overset{(m)}{f}(z/\overset{(m)}{x}).
\end{align}
The above ansatz can significantly simplify both the equation of motion (EOM) and the boundary conditions. First, a more general ansatz, such as \(\overset{(m)}{\Phi}= \overset{(m)}{f_g}(z, \overset{(m)}{x})\), leads to a partial differential equation from the EOM, which is difficult to solve. In contrast, the ansatz (\ref{app2: bulk scalar}) yields a simpler differential equation:
  \begin{align} \label{app2: EOM bulk scalar}
z \left(z^2+1\right)  \overset{(m)}{f}{}''(z)+\left(2 z^2-3\right)  \overset{(m)}{f}{}'(z)=0,
\end{align}
which can be easily solved as:
\begin{align} \label{app2: bulk scalar1}
 \overset{(m)}{f}(z)=\overset{(m)}{c}_1+\overset{(m)}{c}_2\Big( \frac{2 }{3}
-\frac{ \left(3 z^2+2\right)}{3 \left(z^2+1\right)^{3/2}}\Big),
 \end{align}
 where $\overset{(m)}{c}_i$ are constants.  Second, for the general ansatz $\overset{(m)}{\Phi}= \overset{(m)}{f_g}(z, \overset{(m)}{x})$, the JC (\ref{app2: JC1 scalar bulk}) imposes constraints on $\overset{(m)}{f_g}(z, -\sinh(\rho)z)$, which complicates our analysis. However, with the special ansatz (\ref{app2: bulk scalar}), we obtain much simpler constraints on the Net-brane:
\begin{align}\label{app2: BC bulk scalar}
 \overset{(m)}{f}\big(-\text{csch}(\rho )\big)=\overset{(n)}{f}\big(-\text{csch}(\rho )\big), \ \ \  \sum_{m=1}^p \overset{(m)}{f}{{'}}\big(-\text{csch}(\rho )\big)=0.
\end{align}
Substituting the solution (\ref{app2: bulk scalar1}) into the above boundary conditions, we derive
  \begin{align} \label{app2: bulk scalar3}
\overset{(m)}{c}_1+\frac{1}{3} \overset{(m)}{c}_2 \left(2-\tanh (\rho ) \left(\text{sech}^2(\rho )+2\right)\right)=\text{c}, \ \ \ \ \ \   \sum_m^p\overset{(m)}{c}_2=0, 
\end{align}
where $c$ is a constant independent of $m$.   
 
According to AdS/CFT, as we approach the boundary of AdS, the massless bulk scalar behaves like:
\begin{align}\label{app2: dual operator}
\Phi = \phi+ \frac{z^4}{4}  \langle O\rangle+...,
\end{align}
where $\phi$ is the background scalar field on networks and $O$ is the dual operator. Comparing (\ref{app2: bulk scalar},\ref{app2: bulk scalar1}) with (\ref{app2: dual operator}), we get 
 \begin{align} \label{app2: bulk scalar2}
\overset{(m)}{\phi}=\overset{(m)}{c}_1, \ \ \ \langle \overset{(m)}{O} \rangle=\frac{\overset{(m)}{c}_2}{\overset{(m)}{x}{}^4},
 \end{align}
 which takes the expected form of a one-point function for a scalar operator. 
Please note that the first term of (\ref{app2: bulk scalar3}) imposes a constraint on the source term $\overset{(m)}{\phi}=\overset{(m)}{c}_1$. This constraint arises from the continuity condition of the bulk scalar, as indicated by the first term of (\ref{app2: BC bulk scalar}) on the Net-brane, which is a natural requirement. It corresponds to the scenario of a continuous induced metric at the network node, which also imposes constraints on the source (metric) of the stress tensor. 
We emphasize that this constraint is weak and does not completely determine the source terms.

 From (\ref{app2: bulk scalar3}) and (\ref{app2: bulk scalar2}), we get at the renormalized node $N_{\epsilon} (\overset{(m)}{x}=\epsilon)$:
\begin{align} \label{app2: JC1 O}
\sum_{m=1}^p  \langle \overset{(m)}{O} \rangle |_{N_{\epsilon}}=0.
\end{align}
It implies a discontinuous scalar operator $O$ at the node, which corresponds to the case of stress tensor $\sum_{m} \overset{(m)}{T}_{na}|_{\text{node}}=0$. 
Now we finish the discussions of the holographic one-point function of the scalar operator.

\section{Holographic two-point functions in general network}
\label{app for two point function}

 This appendix examines the two-point function of scalar operators in AdS/NCFT for general networks. To simplify our analysis, we focus on the massive scalar field in multi-Poincaré AdS glued by tensionless Net-branes. It's important to note that the Poincaré AdS is not dual to the vacuum state of NCFTs in general networks. However, we selected it because it significantly simplifies the calculations. Additionally, this choice is related to the holographic shortest path problem discussed in Section 5.

In AdS/CFT \cite{Witten:1998qj}, the bulk scalar field $\Phi$ can be obtained from the boundary source $\phi$ via the boundary-to-bulk propagator $K_0$
\begin{align} \label{app C: K in AdSCFT}
    \Phi(z,\textbf{x})=\int d^{d}\mathbf{x}' K_0(z,\mathbf{x},\mathbf{x}') \phi(\mathbf{x}') ,~ K_0(z,\mathbf{x},\mathbf{x}')=c' \ \frac{z^{\Delta}}{(z^2+|\mathbf{x}-\mathbf{x}'|^2)^{\Delta}},
\end{align}
where $\mathbf{x}'$ are coordinates on the AdS boundary $z=0$, $\Delta$ is the conformal dimension and $c'$ is a constant. 

In AdS/NCFT, we consider the following Euclidean action of bulk scalar fields
\begin{align} \label{app C: action}
    I=\frac{1}{2}\sum_{m}\int_{B_{m}}d^{d+1}x \left(\nabla_{\mu}\overset{(m)}{\Phi}\nabla^{\mu}\overset{(m)}{\Phi}+M^{2}\overset{(m)}{\Phi}{}^{2} \right).
\end{align}
The scalar $\overset{(m)}{\Phi}$ obeys the equation of motion (EOM) in the bulk
\begin{align} \label{app C: Phi EOM}
    z^{d+1}\partial_{z}(z^{1-d}\partial_{z}\overset{(m)}{\Phi})+z^{2}\delta^{ij}\partial_{i}\partial_{j}\overset{(m)}{\Phi}-M^{2}\overset{(m)}{\Phi}=0,
\end{align}
and the junction condition (JC) (\ref{sect2: bdy dI matter}) on the Net-brane
\begin{align} \label{app C: JC}
    \overset{(q)}{\Phi}|_{NB}=\overset{(m)}{\Phi}|_{NB},~\sum_{m}\partial_{\hat{n}}\overset{(m)}{\Phi}|_{NB}=0.
\end{align}
Here, \(B_{m}\) and \(B_{q}\) refer to the branches linked to the chosen Net-brane (NB), and \(\hat{n}\) indicates the normal direction from the branch to the Net-brane. 
For the case of tensionless branes, \(\hat{n}\) aligns with the \(\overset{(m)}{x}\) direction on the branch \(B_m\). 
The boundary condition on the network edge \(E_{m}\) is given by:
\begin{align}\label{app C: Phi BC}
    \overset{(m)}{\Phi}|_{E_{m}}=\epsilon^{d-\Delta}\overset{(m)}{\phi},
\end{align}
where $z=\epsilon$ represents the UV cutoff of the AdS boundary.

Similar to (\ref{app C: K in AdSCFT}), the field $\overset{(m)}{\Phi}$ in the bulk branch $B_{m}$ is obtained by integrating the boundary source $\phi$ over all edges $E_{q}$ using the boundary-to-bulk propagator $K$:
\begin{align} \label{app C: Phi from phi}
    \overset{(m)}{\Phi}=\Phi(z,\overset{(m)}{\mathbf{x}})=\sum_{q}\int_{E_{q}} d^{d}\overset{(q)}{\mathbf{x}}{}' K(z,\overset{(m)}{\mathbf{x}},\overset{(q)}{\mathbf{x}}{}') \phi(\overset{(q)}{\mathbf{x}}{}'),
\end{align}
where the propagator \(K\) adheres to the same EOM (\ref{app C: Phi EOM}) as the bulk scalar fields. The boundary conditions (BC) and junction conditions (JC) for the propagator on the Net-brane and the AdS boundary can be deduced from equations (\ref{app C: JC} and \ref{app C: Phi BC}) that apply to bulk scalars. Specifically, we have:
\begin{align} \label{app C: JC K}
   K(z,\overset{(m)}{\mathbf{x}},\overset{(q)}{\mathbf{x}}{}')|_{\overset{(m)}{x}=0}= K(z,\overset{(n)}{\mathbf{x}},\overset{(q)}{\mathbf{x}}{}')|_{\overset{(n)}{x}=0},~\sum_{m}\partial_{\hat{n}}K(z,\overset{(m)}{\mathbf{x}},\overset{(q)}{\mathbf{x}}{}')|_{\overset{(m)}{x}=0}=0,
\end{align}
and
\begin{align} \label{app C: K delta BC}
  \lim_{z\to \epsilon \to 0}K(z,\overset{(m)}{\mathbf{x}},\overset{(q)}{\mathbf{x}}{}')=\epsilon^{d-\Delta}\delta_{m,q}\delta^d(\overset{(m)}{\mathbf{x}}-\overset{(q)}{\mathbf{x}}{}'),
\end{align}
where $\overset{(m)}{\mathbf{x}}=(\overset{(m)}{x},\overset{(m)}{\mathbf{y}})$ with $\overset{(m)}{x}=0$ denoting the node and tensionless Net-branes. 

Remarkably, the boundary-to-bulk propagator $K$ takes on an illuminating form 
\begin{align} \label{app C: K in AdSNCFT}
    K(z,\overset{(m)}{\mathbf{x}},\overset{(q)}{\mathbf{x}}{}') =c' \sum_{\text{all path }\mathcal{P}_{i}}\frac{(\Pi_{a \in\mathcal{P}_{i}}c_{a})z^{\Delta}}{\left(z^{2}+|\overset{(m)}{\mathbf{x}}-\overset{(q)}{\mathbf{x}}{}'|_{i}^{2}\right)^{\Delta}},
\end{align}
where $\mathcal{P}_{i}$ represents the shortest path within every set of isomorphic boundary paths connecting two boundary points $\overset{(q)}{\mathbf{x}}{}'$ and $\overset{(m)}{\mathbf{x}}$, and may include loops along edges. By ``isomorphism", we mean boundary paths that can be continuously transformed into one another without leaving the nodes or crossing into new nodes. 
There are infinite isomorphic boundary paths connecting two points and we choose the one with shortest length as $\mathcal{P}_{i}$ and denote its length by $|\overset{(m)}{\mathbf{x}}-\overset{(q)}{\mathbf{x}}{}'|_{i}$. For two points on the same edge and the paths without touching any nodes, \( |\overset{(m)}{\mathbf{x}} - \overset{(q)}{\mathbf{x}}{}'|_{i} \) is simply the length of the straight line connecting the two points. 
The path weight \(c_a\) is calculated from the reflection coefficient \(c_{r}\) and the transmission coefficient \(c_{t}\):
\begin{align}\label{app C: crct}
  c_{r}=\frac{2-p}{p},~c_{t}=\frac{2}{p},
\end{align}
where \(p\) is the number of edges at a node, which may vary between nodes. When a boundary
path passes through a node, it receives a factor of \(c_t\). When reflected at a node, it receives a factor of \(c_r\). If there is no reflection or transmission in $\mathcal{P}_{i}$, we take $\Pi_{a \in\mathcal{P}_{i}}c_{a}=1$. This framework aligns with the physical picture of wave reflection and transmission in networks, as is discussed in (\ref{sect3: probability conservation}-\ref{sect3: cr ct meaning 3}) of Sec. 3. 

We want to emphasize that the boundary-to-bulk propagator \( K \) (\ref{app C: K in AdSNCFT}) and the bulk scalar \(  \overset{(m)}{\Phi} \)  (\ref{app C: Phi from phi}) satisfy the equations of motion (EOM) in the bulk, the junction conditions on the Net-brane, and the boundary conditions at the AdS boundary. We will verify these statements below.

Note that the propagator (\ref{app C: K in AdSNCFT}) in AdS/NCFT has the same dependence on $\overset{(m)}{\mathbf{x}}$ and $z$ as the propagator (\ref{app C: K in AdSCFT}) in AdS/CFT. It is simply the weighted sum of the propagator in AdS/CFT. Consequently, it automatically satisfies the EOM (\ref{app C: Phi EOM}). For the same reason, it obeys the BC (\ref{app C: K delta BC}) on the AdS boundary. From (\ref{app C: K in AdSNCFT}), we get 
\begin{align} \label{app C: check delta}
     \lim_{z\to \epsilon \to 0}K(z,\overset{(m)}{\mathbf{x}},\overset{(q)}{\mathbf{x}}{}') =\epsilon^{d-\Delta}  \sum_{\text{all path }\mathcal{P}_{i}}(\Pi_{a \in\mathcal{P}_{i}}c_{a})\delta^d(\overset{(m)}{\mathbf{x}}-\overset{(q)}{\mathbf{x}}{}')_i=\epsilon^{d-\Delta}\delta_{m,q} \delta^d(\overset{(m)}{\mathbf{x}}-\overset{(q)}{\mathbf{x}}{}'). 
\end{align}
To derive the final equality, we utilized the fact that when the two points are located on different edges or when the path connecting them includes any node, the distance \( |\overset{(m)}{\mathbf{x}} - \overset{(q)}{\mathbf{x}}{}'|_{i} \) is greater than zero, leading to \( \delta^d ( \overset{(m)}{\mathbf{x}} - \overset{(q)}{\mathbf{x}}{}' )_{i} = 0 \). Thus, the corresponding delta function contributes to the final result only when the two points lie on the same edge and the path $\mathcal{P}_{i}$ connecting them contains no nodes. It corresponds to the case $\Pi_{a \in\mathcal{P}_{i}}c_{a}=1$ and yields the last equality of (\ref{app C: check delta}).

The physical picture of wave reflection and transmission indicates that the propagator (\ref{app C: K in AdSNCFT}) automatically obeys the JC (\ref{app C: JC K}). Let us explain the main ideas here and provide solid proof at the end of the appendix. Consider a selected node connected by $p$ edges $E_{m}$, to which $p$ propagators (\ref{app C: K in AdSNCFT}) approach as $\overset{(m)}{x} \to 0$. 
These propagators encompass all possible scenarios for the free propagation, reflection, and transmission of waves near the node. We can categorize them into a set of basic elements, where each element includes one wave-free propagation and reflection at one edge, along with \( (p-1) \) transmissions at the remaining edges. As discussed (\ref{sect3: cr ct meaning 1}-\ref{sect3: cr ct meaning 3}) of Sec. 3, each of these basic elements complies with the junction condition. Therefore, the entire set formed by these elements also adheres to the junction condition.

Substituting (\ref{app C: Phi from phi}) with the propagator (\ref{app C: K in AdSNCFT}) into the action (\ref{app C: action}) and using the BC (\ref{app C: K delta BC}), we obtain 
\begin{align} \label{app C: action two point}
    I(\phi)=\frac{c' \Delta}{2}\sum_{m}\sum_{q}\int_{E_{m}}d^{d}\overset{(m)}{\mathbf{x}}\int_{E_{q}}d^{d}\overset{(q)}{\mathbf{x}}{}' \phi_{0}(\overset{(m)}{\mathbf{x}})\phi_{0}(\overset{(q)}{\mathbf{x}}{}')\sum_{\text{all path }\mathcal{P}_{i}}\frac{\Pi_{a \in\mathcal{P}_{i}}c_{a}}{|\overset{(m)}{\mathbf{x}}-\overset{(q)}{\mathbf{x}}{}'|_{i}^{2\Delta}},
\end{align}
which yields the holographic two-point functions of scalar operators
\begin{align}  \label{app C: holo two point}
    \langle O(\overset{(m)}{\mathbf{x}})O(\overset{(q)}{\mathbf{x}}{}') \rangle=\frac{\delta^{2}I}{\delta\phi(\overset{(m)}{\mathbf{x}})\delta \phi(\overset{(q)}{\mathbf{x}}{}')}=c'\Delta\sum_{\text{all path }\mathcal{P}_{i}}\frac{\Pi_{a \in\mathcal{P}_{i}}c_{a}}{|\overset{(m)}{\mathbf{x}}-\overset{(q)}{\mathbf{x}}{}'|_{i}^{2\Delta}}.
\end{align} \\

  \begin{figure}[htbp]
  \centering
\includegraphics[width=0.4\textwidth]{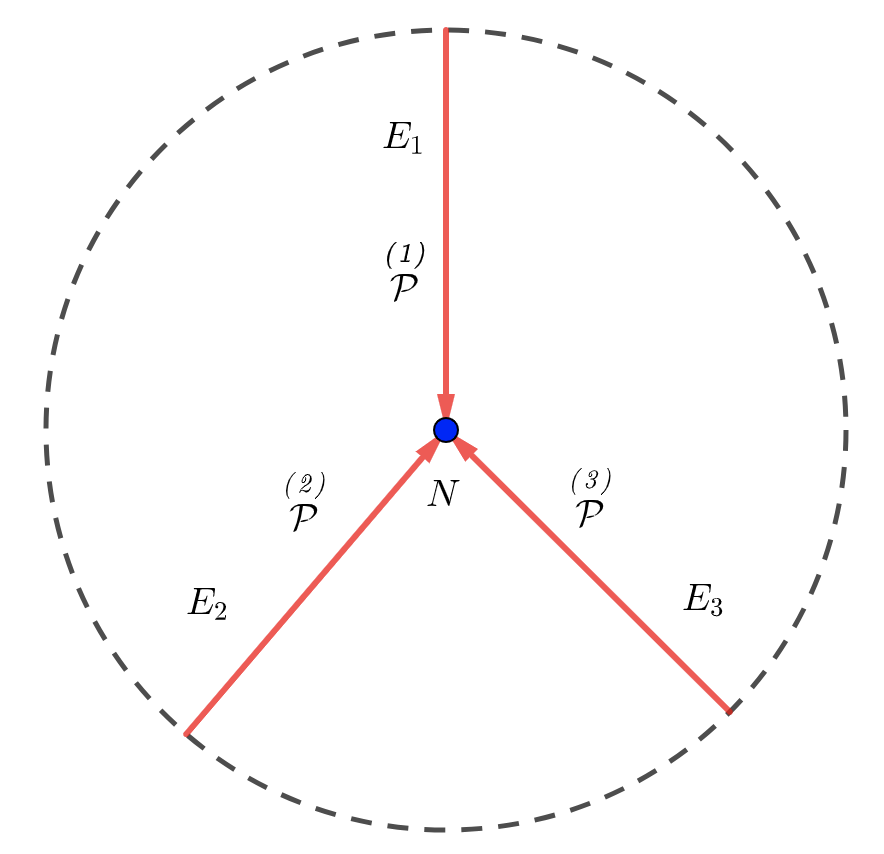} \includegraphics[width=0.4\textwidth]{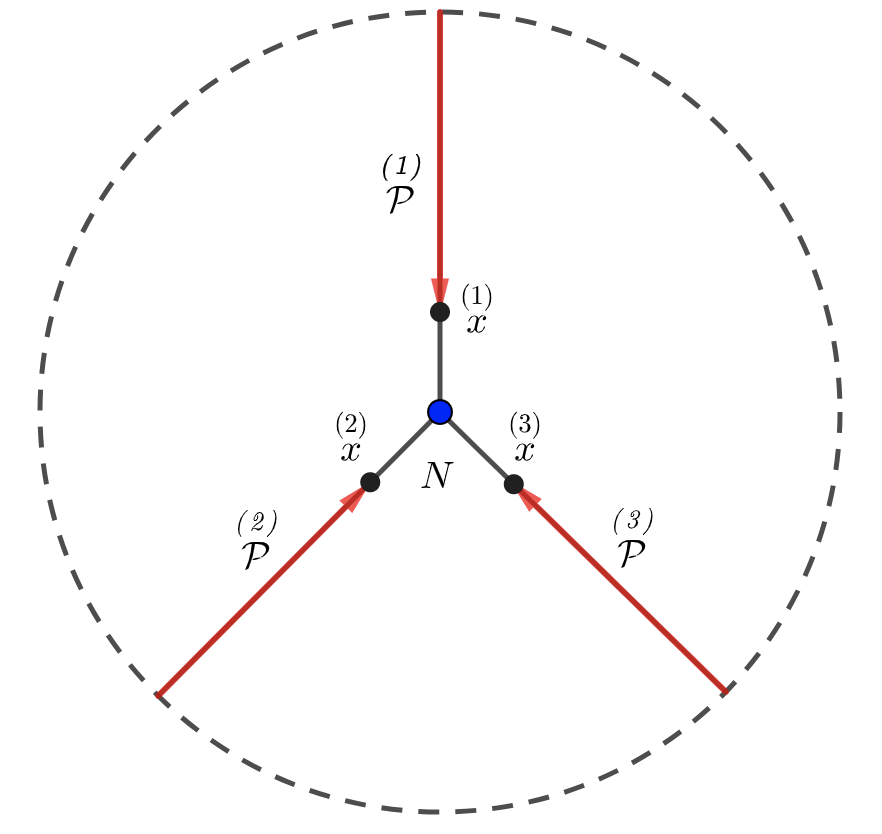}
 \caption{A local region (dotted circle) near a node (blue point) in networks. In the left figure, we consider the paths from a general outside point $\overset{(q)}{\mathbf{x}}{}'$ to the node $\overset{(m)}{x}=0$. Denote the path pointing from the edge $E_m$ to the node by $\overset{(m)}{\mathcal{P}}$ (red vector). In the right figure, we consider the path set $\mathcal{P}(\overset{(1)}{\mathbf{x}}, \overset{(q)}{\mathbf{x}}{}')$ from an outside point $\overset{(q)}{\mathbf{x}}{}'$ to the point  $\overset{(m)}{\mathbf{x}}$ near the node. Take the point $\overset{(1)}{\mathbf{x}}$ as an example, the path set $\mathcal{P}(\overset{(1)}{\mathbf{x}}, \overset{(q)}{\mathbf{x}}{}')$ includes not only $\overset{(1)}{\mathcal{P}}$ (red vector) but also one additional set of reflection paths, $\overset{(1)}{\mathcal{P}} \cdot\mathcal{P}_{1\to N\to 1}$, and two additional sets of transmission paths, $\overset{(2)}{\mathcal{P}}\cdot \mathcal{P}_{2\to N\to 1}$ and $\overset{(3)}{\mathcal{P}} \cdot \mathcal{P}_{3\to N \to1}$.}
 \label{paths}
\end{figure}

To end this appendix, let us prove that the propagator (\ref{app C: K in AdSNCFT}) obeys the JC (\ref{app C: JC K}).

Let us classify the paths from a general point \(\overset{(q)}{\mathbf{x}}{}'\) to the point \(\overset{(m)}{\mathbf{x}}\) located near the node \(\overset{(m)}{x} = 0\). We will adopt a two-step approach, as illustrated in Fig. \ref{paths}. 
First, we examine a local region around the node \(\overset{(m)}{x} = 0\), and include all the paths extending from the edge \(E_m\) to this node in the set 
\(\overset{(m)}{\mathcal{P}}\). See Fig. \ref{paths} (Left). It is important to note that each path in \(\overset{(m)}{\mathcal{P}}\) may intersect with other edges during the intermediate stages of its trajectory. Our focus is on the trailing end of the path when defining $\overset{(m)}{\mathcal{P}}$. In the second step, as shown in Fig. \ref{paths} (Right), we consider the set of paths \(\mathcal{P}\) from an outside point \(\overset{(q)}{\mathbf{x}}{}'\) to the point \(\overset{(m)}{\mathbf{x}}\) near the node. For example, let us take the point \(\overset{(1)}{\mathbf{x}}\). The path set $\mathcal{P}(\overset{(1)}{\mathbf{x}}, \overset{(q)}{\mathbf{x}}{}')$ includes not only $\overset{(1)}{\mathcal{P}}(\overset{(1)}{\mathbf{x}}, \overset{(q)}{\mathbf{x}}{}')$ but also an additional set of reflection paths, $\overset{(1)}{\mathcal{P}} \cdot \mathcal{P}_{1\to N\to 1}$, as well as $(p-1)$ sets of transmission paths, $\overset{(n)}{\mathcal{P}} \cdot \mathcal{P}_{n\to N\to 1}$ for $n \ne 1$. No other paths exist since any paths involving more than one reflection and transmission have already been accounted for within \(\overset{(m)}{\mathcal{P}}\). 

In summary, the path set \(\mathcal{P}\) from a general point \(\overset{(q)}{\mathbf{x}}{}'\) to the point \(\overset{(m)}{\mathbf{x}}\), which is located near the node, can be classified as follows:
\begin{align}  \label{app C: path set}
\mathcal{P}(\overset{(m)}{\mathbf{x}}, \overset{(q)}{\mathbf{x}}{}')=\{  \overset{(m)}{\mathcal{P}},\ \overset{(m)}{\mathcal{P}} \cdot \mathcal{P}_{m\to N\to m},\ \overset{(n)}{\mathcal{P}} \cdot \mathcal{P}_{n\to N\to m} \}
\end{align}
where $n\ne m$, and the edges $E_m, E_n$ are connected through the same node $N$. Here, \(\mathcal{P}_{m \to N \to m}\) represents the reflection along the edge \(E_m\), while \(\mathcal{P}_{n \to N \to m}\) denotes the transmission from edge \(E_n\) to edge \(E_m\).  The squared distance of the path in \(\overset{(m)}{\mathcal{P}}\), which connects the point \(\overset{(q)}{\mathbf{x}}{}'\) and the node \(\overset{(m)}{x} = 0\), is denoted by 
\begin{align}  \label{app C: path node distance}
|\overset{(m)}{\mathcal{P}}(\overset{(m)}{\mathbf{x}},\overset{(q)}{\mathbf{x}}{}')|^2_{\overset{(m)}{x}=0}=\overset{(m)}{L}{}^2+(\overset{(m)}{\mathbf{y}}-\overset{(q)}{\mathbf{y}}{}')^{2}.
\end{align}
Here, $\overset{(m)}{L}$ represents the length associated with the path in $\overset{(m)}{\mathcal{P}}$. Although the lengths may vary among different paths in $\overset{(m)}{\mathcal{P}}$, they are collectively denoted by $\overset{(m)}{L}$ for simplicity. As illustrated in Fig. \ref{paths} (Right), the corresponding squared distances of the paths in (\ref{app C: path set}) are given by:
\begin{align}  \label{app C: path set distance}
|\mathcal{P}(\overset{(m)}{\mathbf{x}},\overset{(q)}{\mathbf{x}}{}')|=\{|\overset{(m)}{\mathbf{x}}-\overset{(q)}{\mathbf{x}}{}'|_{i}^2\}=&\{(\overset{(m)}{L}-\overset{(m)}{x})^{2}+(\overset{(m)}{\mathbf{y}}-\overset{(q)}{\mathbf{y}}{}')^{2},  \ (\overset{(m)}{L}+\overset{(m)}{x})^{2}+(\overset{(m)}{\mathbf{y}}-\overset{(q)}{\mathbf{y}}{}')^{2},  \nonumber\\
&(\overset{(n)}{L}+\overset{(m)}{x})^{2}+(\overset{(m)}{\mathbf{y}}-\overset{(q)}{\mathbf{y}}{}')^{2} \}.
\end{align}

By using equations (\ref{app C: path set}) and (\ref{app C: path set distance}), we can rewrite the propagator (\ref{app C: K in AdSNCFT}) as follows:
\begin{align} \label{app C: K path}
    K(z,\overset{(m)}{\mathbf{x}},\overset{(q)}{\mathbf{x}}{}') &=c' \sum_{\overset{(m)}{\mathcal{P}}} \Big[\frac{(\Pi_{\overset{(m)}{\mathcal{P}}} c_{a})z^{\Delta}}{\left(z^{2}+(\overset{(m)}{L}-\overset{(m)}{x})^{2}+(\overset{(m)}{\mathbf{y}}-\overset{(q)}{\mathbf{y}}{}')^{2}\right)^{\Delta}}+\frac{c_r\ (\Pi_{\overset{(m)}{\mathcal{P}}} c_{a})z^{\Delta}}{\left(z^{2}+(\overset{(m)}{L}+\overset{(m)}{x})^{2}+(\overset{(m)}{\mathbf{y}}-\overset{(q)}{\mathbf{y}}{}')^{2}\right)^{\Delta}}\Big]\nonumber\\
    &+c' \sum_{n\neq m}\sum_{\overset{(n)}{\mathcal{P}}} \frac{c_t\ (\Pi_{\overset{(n)}{\mathcal{P}}} c_{a})z^{\Delta}}{\left(z^{2}+(\overset{(n)}{L}+\overset{(m)}{x})^{2}+(\overset{(m)}{\mathbf{y}}-\overset{(q)}{\mathbf{y}}{}')^{2}\right)^{\Delta}}
\end{align}
where the first, second, and last terms correspond to free propagation, reflection, and transmission, respectively. Note that the subscript $\overset{(m)}{\mathcal{P}}$ indicates that both the summation and the product are associated with it; the explicit form of $\sum_{\overset{(m)}{\mathcal{P}}}$ and $\Pi_{\overset{(m)}{\mathcal{P}}}$ follows (\ref{app C: K in AdSNCFT}).
One can directly verify that the propagator (\ref{app C: K path}) satisfies the JC (\ref{app C: JC K}). To demonstrate this, it is helpful to combine the \( p \) propagators near the node at \(\overset{(m)}{x} = 0\). Without loss of generality, we can choose a basic propagation process among the \( p \) propagators involving free propagation and reflection at edge \( E_1 \), along with \( (p-1) \) transmissions from \( E_1 \) to the other edges \( E_n \) where \( n = 2, 3, \ldots, p \):
\begin{align} \label{app C: K path basic 1}
& \hat{K}(z,\overset{(1)}{\mathbf{x}},\overset{(q)}{\mathbf{x}}{}')=c' \sum_{\overset{(1)}{\mathcal{P}}} \Big[\frac{(\Pi_{\overset{(1)}{\mathcal{P}}} c_{a})z^{\Delta}}{\left(z^{2}+(\overset{(1)}{L}-\overset{(1)}{x})^{2}+(\overset{(1)}{\mathbf{y}}-\overset{(q)}{\mathbf{y}}{}')^{2}\right)^{\Delta}}+\frac{c_r\ (\Pi_{\overset{(1)}{\mathcal{P}}} c_{a})z^{\Delta}}{\left(z^{2}+(\overset{(1)}{L}+\overset{(1)}{x})^{2}+(\overset{(1)}{\mathbf{y}}-\overset{(q)}{\mathbf{y}}{}')^{2}\right)^{\Delta}}\Big],\\
 & \hat{K}(z,\overset{(n)}{\mathbf{x}},\overset{(q)}{\mathbf{x}}{}')=c' \sum_{\overset{(1)}{\mathcal{P}}} \frac{c_t\ (\Pi_{\overset{(1)}{\mathcal{P}}} c_{a})z^{\Delta}}{\left(z^{2}+(\overset{(1)}{L}+\overset{(n)}{x})^{2}+(\overset{(n)}{\mathbf{y}}-\overset{(q)}{\mathbf{y}}{}')^{2}\right)^{\Delta}},  \label{app C: K path basic 2}
\end{align}
where the summation includes only the paths in the set $\overset{(1)}{\mathcal{P}}$. Similarly, we can choose the basic propagation process with contributions from other sets \(\overset{(n)}{\mathcal{P}}\). In total, these formulations yield the complete \( p \)-propagators near the node.

One can easily check that the basic elements (\ref{app C: K path basic 1},\ref{app C: K path basic 2}) satisfy the JC (\ref{app C: JC K}). 
At the identical points \(\overset{(1)}{\mathbf{y}} = \overset{(n)}{\mathbf{y}}\) of the Net-brane where \(\overset{(1)}{x} = \overset{(n)}{x} = 0\), we have:
\begin{align} \label{app C: JC proof 1}
& \hat{K}(z,\overset{(1)}{\mathbf{x}},\overset{(q)}{\mathbf{x}}{}')|_{\overset{(1)}{x}=0}- \hat{K}(z,\overset{(n)}{\mathbf{x}},\overset{(q)}{\mathbf{x}}{}')|_{\overset{(n)}{x}=0}\nonumber\\
& =c' \sum_{\overset{(1)}{\mathcal{P}}} \frac{\ (\Pi_{\overset{(1)}{\mathcal{P}}} c_{a})z^{\Delta}}{\left(z^{2}+\overset{(1)}{L}{}^{2}+(\overset{(1)}{\mathbf{y}}-\overset{(q)}{\mathbf{y}}{}')^{2}\right)^{\Delta}} \Big( 1+ c_r-c_t\Big)=0,
\end{align}
and 
\begin{align} \label{app C: JC proof 2}
& \partial_{\overset{(1)}{x}}\hat{K}(z,\overset{(1)}{\mathbf{x}},\overset{(q)}{\mathbf{x}}{}')|_{\overset{(1)}{x}=0}+\sum_{n=2}^p\partial_{\overset{(n)}{x} }\hat{K}(z,\overset{(n)}{\mathbf{x}},\overset{(q)}{\mathbf{x}}{}')|_{\overset{(n)}{x}=0}\nonumber\\
& =c' \sum_{\overset{(1)}{\mathcal{P}}} \frac{\  2\Delta \overset{(1)}{L} (\Pi_{\overset{(1)}{\mathcal{P}}} c_{a})z^{\Delta}}{\left(z^{2}+\overset{(1)}{L}{}^{2}+(\overset{(1)}{\mathbf{y}}-\overset{(q)}{\mathbf{y}}{}')^{2}\right)^{\Delta+1}} \Big( 1-c_r-(p-1)c_t\Big)=0,
\end{align}
where we have used \( c_r = (2 - p)/p \) and \( c_t = 2/p \) above. Since the basic elements (\ref{app C: K path basic 1}, \ref{app C: K path basic 2}) obey the junction condition, the total propagator (\ref{app C: K path}) also satisfies the criterion. Thus, we conclude our proof.




\end{document}